\documentclass[useAMS,usenatbib]{mn2e}
\usepackage{graphicx}
\usepackage{epsfig}
\usepackage{rotating}
\usepackage{amssymb}
\usepackage{color}

\usepackage{fancyhdr}
\usepackage{textcomp}
\usepackage{longtable}
\usepackage{tabularx}
\usepackage[abs]{overpic}
\usepackage{tocloft}

\def\um{$\textrm{$\mu$m}\ $}

\def\arcsec{\hbox{$^{\prime\prime}$}}                       
\def  \cmsq     {\ifmmode {\rm cm}^{-2} \else cm$^{-2}$\fi}
\def  \ergs     {\ifmmode {\rm erg\,s}^{-1} \else erg s$^{-1}$\fi}
\def  \ergcms   {\ifmmode {\rm erg\,cm}^{-2}\,{\rm s}^{-1}
                        \else erg\,cm$^{-2}$\,s$^{-1}$\fi}
\def \lhard  {\ifmmode {\rm L_{2-10keV}} \else ${\rm L_{2-10keV}}$\fi}
\def \lir  {\ifmmode {\rm L_{IR}} \else ${\rm L_{IR}}$\fi}
\def \nh  {\ifmmode {\rm N_{H}} \else ${\rm N_{H}}$\fi}
\def \Msun {\ifmmode M_{\odot} \else $M_{\odot}$\fi}
\def \Lsun {\ifmmode L_{\odot} \else $L_{\odot}$\fi}

\def \lephare	{{\it Le PHARE \,}}	
\def   \sun	{\odot}	
\def \rvir	 {\ifmmode R_{200} \else $\rm R_{200}$\fi}

\title[Lack of star formation gradients in groups]{The lack of star formation
  gradients in galaxy groups up to $z\sim1.6$}
\author[F. Ziparo et al.]{F. Ziparo$^{1,2}$\thanks{E-mail:
fziparo@star.sr.bham.ac.uk}, P. Popesso$^{1}$, A. Biviano$^{3}$, A. Finoguenov$^{1, 4, 5}$, S. Wuyts$^{1}$, D. Wilman$^{1}$, \newauthor
M.~Salvato${^1}$, M.~Tanaka$^{6}$, O.~Ilbert$^{7}$,  K.~Nandra$^{^1}$, D.~Lutz$^{1}$, D.~Elbaz $^{8}$, M.~Dickinson$^{9}$, 
 \newauthor   B.~Altieri$^{10}$, H. Aussel$^{8}$, S. Berta$^{1}$, A. Cimatti$^{11}$, D. Fadda$^{12}$, R. Genzel${^1}$, E. Le Flo'ch$^{8}$,  
 \newauthor  B. Magnelli$^{13}$, R. Nordon$^{14}$, A. Poglitsch$^{1}$, F. Pozzi$^{11}$ M. Sanchez Portal$^{10}$, L. Tacconi${^1}$  
 \newauthor F. E. Bauer$^{15,16}$, W. N. Brandt$^{17}$, N. Cappelluti$^{18, 5}$, M. C. Cooper$^{19}$, J. S. Mulchaey$^{20}$ \\
$^{1}$Max-Planck-Institut f\"{u}r extraterrestrische Physik, Giessenbachstra\ss e 1, 85748 Garching bei M\"{u}nchen, Germany\\
$^{2}$School of Physics and Astronomy, University of Birmingham, Edgbaston, Birmingham B15 2TT, UK\\
$^{3}$INAF/Osservatorio Astronomico di Trieste, via G. B. Tiepolo 11, 34143 Trieste, Italy\\
$^{4}$Department of Physics, University of Helsinki, Gustaf H\"allstr\"omin katu 2a, 00014 Helsinki, Finland \\
$^{5}$University of Maryland Baltimore County, 1000 Hilltop circle, Baltimore, MD 21250, USA \\
$^{6}$National Astronomical Observatory of Japan, 2-21-1 Osawa, Mitaka, Tokyo 181-8588, JAPAN \\
$^{7}$Institute for Astronomy 2680 Woodlawn Drive Honolulu, HI 96822-1897, USA \\
$^{8}$Laboratoire AIM, CEA/DSM-CNRS-Universit{\'e} Paris Diderot, IRFU/Service d'Astrophysique,  B\^at.709, CEA-Saclay, 91191 Gif-sur-Yvette \\
Cedex, France.\\
$^{9}$National Optical Astronomy Observatory, 950 North Cherry Avenue, Tucson, AZ 85719, USA\\
$^{10}$Herschel Science Centre, European Space Astronomy Centre, ESA, Villanueva de la Ca\~nada, 28691 Madrid, Spain\\
$^{11}$Dipartimento di Astronomia, Universit{\`a} di Bologna, Via Ranzani 1, 40127 Bologna, Italy.\\
$^{12}$NASA Herschel Science Center, Caltech 100-22, Pasadena, CA 91125, USA\\
$^{13}$Argelander-Institut f\"ur Astronomie, Universität Bonn, Auf dem H\"ugel 71, 53121 Bonn, Germany \\
$^{14}$School of Physics and Astronomy, The Raymond and Beverly Sackler Faculty of Exact Sciences, Tel Aviv University, Tel Aviv 69978, Israel \\
$^{15}$Instituto de Astrof\'{\i}sica, Facultad de F\'{i}sica, Pontificia Universidad Catlica de Chile, 306, Santiago 22, Chile \\
$^{16}$Space Science Institute, 4750 Walnut Street, Suite 205, Boulder, Colorado 80301 \\
$^{17}$Department of Astronomy and Astrophysics, 525 Davey Laboratory, The Pennsylvania State University, University Park, PA 16802 \\
$^{18}$INAF-Osservatorio Astronomico di Bologna, Via Ranzani 1 40127 Bologna, Italy \\
$^{19}$Center for Galaxy Evolution, Department of Physics and Astronomy, University of California, Irvine, 4129 Frederick Reines Hall Irvine,\\
CA 92697 USA \\
$^{20}$The Observatoires of the Carnegie Institution of Science, 813 Santa Barbara Street, Pasadena, CA 91101, USA
}

\begin{document}

\date{Accepted 2013 July 2.  Received 2013 June 27; in original form 2013 March 25}

\pagerange{\pageref{firstpage}--\pageref{lastpage}} \pubyear{2012}

\maketitle
\clearpage
\label{firstpage}

\begin{abstract}
In the local Universe, galaxy properties show a strong dependence on environment. In cluster cores,  early-type galaxies dominate, whereas star-forming galaxies are more and more common in the outskirts. At higher redshifts and in somewhat less dense environments (e.g. galaxy groups), the situation is less clear. One open issue is that of whether and how the star formation rate (SFR) of galaxies in groups depends on the distance from the centre of mass. To shed light on this topic, we have built a sample of X-ray selected galaxy groups at $0<z<1.6$ in various blank fields (ECDFS, COSMOS, GOODS).  We use a sample of spectroscopically confirmed group members with stellar mass $\rm M_\star >10^{10.3}\, M_\odot$ in order to have a high spectroscopic completeness. As we use only spectroscopic redshifts, our results are not affected by uncertainties due to projection effects. We use several SFR indicators to link the star formation (SF) activity to the galaxy environment. Taking advantage of the extremely deep mid-infrared Spitzer MIPS and far-infrared $Herschel$\footnotemark[7] PACS observations, we have an accurate, broad-band measure of the SFR for the bulk of the star-forming galaxies.  We use multi-wavelength SED (spectral energy distribution) fitting techniques to estimate the stellar masses of all objects and the SFR of the MIPS and PACS undetected galaxies.  We analyse the dependence of the SF activity, stellar mass and specific SFR on the group-centric distance, up to $z\sim1.6$, for the first time. We do not find any correlation 
between the mean SFR and group-centric distance at any redshift.  We do not observe any strong mass segregation either, in agreement with predictions from simulations.  
Our results suggest that either groups have a much smaller spread in accretion times with respect to the clusters and that the relaxation time is longer than the group crossing time.
\end{abstract}


\footnotetext[7]{$Herschel$ is an ESA space observatory with science instruments provided by European-led Principal Investigator consortia and with important participation from NASA.}


\section{Introduction}

The morphological types of galaxies exhibit differences depending on their large-scale structure environment. More specifically, crowded regions of the nearby Universe have a high fraction of elliptical and lenticular galaxies, while the field is dominated by spirals. A clear manifestation of this is found in galaxy clusters, the densest regions of the Universe, where it has been shown that the fraction of spiral galaxies decreases rapidly from the cluster outskirts towards the dense core \citep{Dressler_1980}. Because spiral galaxies are generally star forming, and early-type galaxies passive, this implies a possible relationship between star formation rate (SFR) and density, and a number of studies have focused on this SFR--density relation \cite[e.g.][]{Balogh2000, Lewis2002, Bai+09, Chung+10, Mahajan2010, Gomez2003}.
In particular,  \cite{Balogh2000} show that the average SFR per galaxy in clusters from the CNOC1 (Canadian Network for Observational Cosmology) survey ($0.19<z<1.55$) is suppressed by almost a factor of two relative to the field, even at distances $\rm \sim 2~ R_{200}$\footnote{$\mathrm{R_\Delta}$ (where $\mathrm{\Delta=500,200}$) is the radius at which the density of a cluster is equal to $\Delta$ times the critical density of the Universe ($\mathrm{\rho_c}$) and $\mathrm{M_\Delta}$ is defined as $\mathrm{M_\Delta=(4 \pi/3) \Delta \rho_c R_{\Delta}^3}$.}. 

Recently, the study of the SFR--density relation has been extended to galaxy groups \cite[e.g.][]{Bai2010,Rasmussen2012,Wetzel2012}. 
There are several reasons why these intermediate environments might be of interest. For example, the majority (50\%-70\%)  of the galaxy population in the local Universe is contained in group-sized haloes of $\rm 10^{12.5}-10^{13.5}$ $M_{\odot}$ \citep{Geller_Huchra1983, Eke2005}.
In addition, in the hierarchical paradigm of structure formation, galaxy groups are the building blocks of massive clusters. Thus, cluster galaxies spend a large fraction of their life in galaxy groups before entering the cluster environment and analysing SFRs in galaxies groups can help assess when, in the hierarchy of halo assembly, the SFR--density relation is established. 

The most recent analyses of gradients of star formation (SF) activity in groups focus on nearby, low redshift systems. 
\cite{Bai2010} find no gradient in the mean star-forming galaxy fraction in a sample of  X-ray detected groups at $ 0.06 < z < 0.1$, observed with {\it{Spitzer}} MIPS 24~$\mu$m. In addition, these groups exhibit a higher star-forming galaxy fraction than the outer region of rich clusters. Exploring the same sample of \cite{Bai2010}, \cite{Rasmussen2012} use deep ultraviolet (UV) observations and detect a SF gradient within $\rm 2~R_{200}$ for galaxies less massive than $\rm 10^{10}~M_\odot$, while they do not find any environmental effect for massive galaxies. A similar conclusion is reached by \cite{Wetzel2012} for a sample of groups in the Sloan Digital Sky Survey (SDSS; \citealt{York2000}). They find that the fraction of galaxies whose SF has been quenched increases towards the halo centre, with a strong trend for the low-mass galaxies.  At somewhat higher redshift, $z\sim 0.3$, \cite{Tran2009} observe an excess of 24~$\mu$m star-forming galaxies with respect to the field in several groups connected to each other and forming a likely ``super-group'', 
or a cluster in formation. It still remains to be established whether this excess is unique to this particular structure or whether it is characteristic of the group environment more generally. 

At much higher redshift ($z{\sim}1.6$), \cite{Tran2010} reveal a very high level of SF activity in one group observed with  {\it{Spitzer}} MIPS 24 $\mu$m. They find a $\sim 2\sigma$ anti-correlation between the level of SF activity and the group-centric distance. They refer to this as a reversal of the relation observed in local clusters. The reason for the reversal is hypothesized to be that, at high redshift, groups show the bulk of the SF in the central massive galaxies that will eventually evolve into early-type galaxies with low SFR by $z\sim 0$, while the group itself will evolve into a massive local cluster.

To shed further light on this topic we have assembled a homogeneously X-ray selected sample of groups at $0 < z \lesssim 1$. We have also considered  a ``super-group'' spectroscopically confirmed at $z\sim 1.6$ by \cite{Kurk2009} and dynamically studied by \cite{popesso2012}. 
Our aim is to understand if the SF gradient observed in the local clusters is in place also in the group regime and at which epoch the gradient is established. For this purpose we use the latest and deepest available {\it{Herschel}} PACS (Photoconducting Array Camera and Spectrometer; \citealt{poglitsch10}) far-infrared surveys, from the PACS Evolutionary Probe (PEP; \citealt{Lutz2010}) and the Great Observatories Origin Deep Survey (GOODS)-Herschel survey (GOODS-H; \citealt{Elbaz2011}). These surveys provide far-infrared observation of the major blank fields, such as the Extended Chandra Deep Field South (ECDFS), the GOODS and the COSMOS (Cosmological Evolution Survey) fields. The use of far-infrared PACS data allows us to overcome the so-called mid-infrared excess problem \cite[due to the uncertain extrapolation of the MIPS 24~$\mu$m flux to extrapolate the bolometric infrared luminosity, $\rm L_{IR}$;][]{elbaz10,nordon10},  and any contamination by active galactic nuclei (AGN) to the optical and mid-infrared emission of the host galaxies \citep{nordon10}. 

This paper is structured as follows: in Section~\ref{dataset} we describe our data set;  Section~\ref{calibration} describes the computation of group membership, velocity dispersion, stellar masses and SFRs; in Section~\ref{sec:completeness_mock_cat} we discuss our approach towards spectroscopic incompleteness; 
Section~\ref{results} shows our main results on the SF gradients in galaxy groups, which are then discussed in Section~\ref{discussion}. Finally, our conclusions are given in Section~\ref{conclusions}.
Throughout our analysis we adopt the AB magnitude system and the following cosmological values: $\mathrm{H_0=70} \mathrm{\ km\ s^{-1}\ Mpc^{-1}}$, $\Omega_\mathrm{M}=0.3$ and $\Omega_\mathrm{\Lambda}=0.7$.

\section{Dataset}
\label{dataset}
The aim of this work is to study the evolution of the SF activity in galaxy groups. We have used X-ray emission to build the group sample. Extended X-ray emitting sources pinpoint groups via the bremsstrahlung radiation of the Intra-Group Medium (IGM). This selects virialized objects \citep[e.g.][]{Forman_Jones1982}, and avoids projection effects which can be problematic with optical selection techniques. 
Optically selected systems which are not X-ray bright tend to be less evolved \cite[e.g.][]{Connelly2012}, and might have more ongoing accretion of galaxies which are not-yet accreted on to the group itself, or even galaxies projected along filaments in the line of sight which are not bound to the group. X-ray selection ensures that a relatively relaxed halo of reasonable mass exists. Once the groups are identified by their X-ray emission, deep multi-wavelength data are required to identify the group members and to determine their properties. Specifically, we require extensive optical photometric and spectroscopic catalogues to identify galaxies and establish their group membership, and Herschel data to determine their SFRs. The necessary combination of data exists in our four fields: ECDFS, COSMOS, GOODS-North and GOODS-South. Throughout our analysis we will use spectroscopic redshifts to define the group membership and study galaxy properties. For calibration purposes we will also make use of photometric redshifts.


\subsection{Extended Chandra Deep Field-South}
\begin{figure*}
\centering
\includegraphics[width=0.48\hsize]{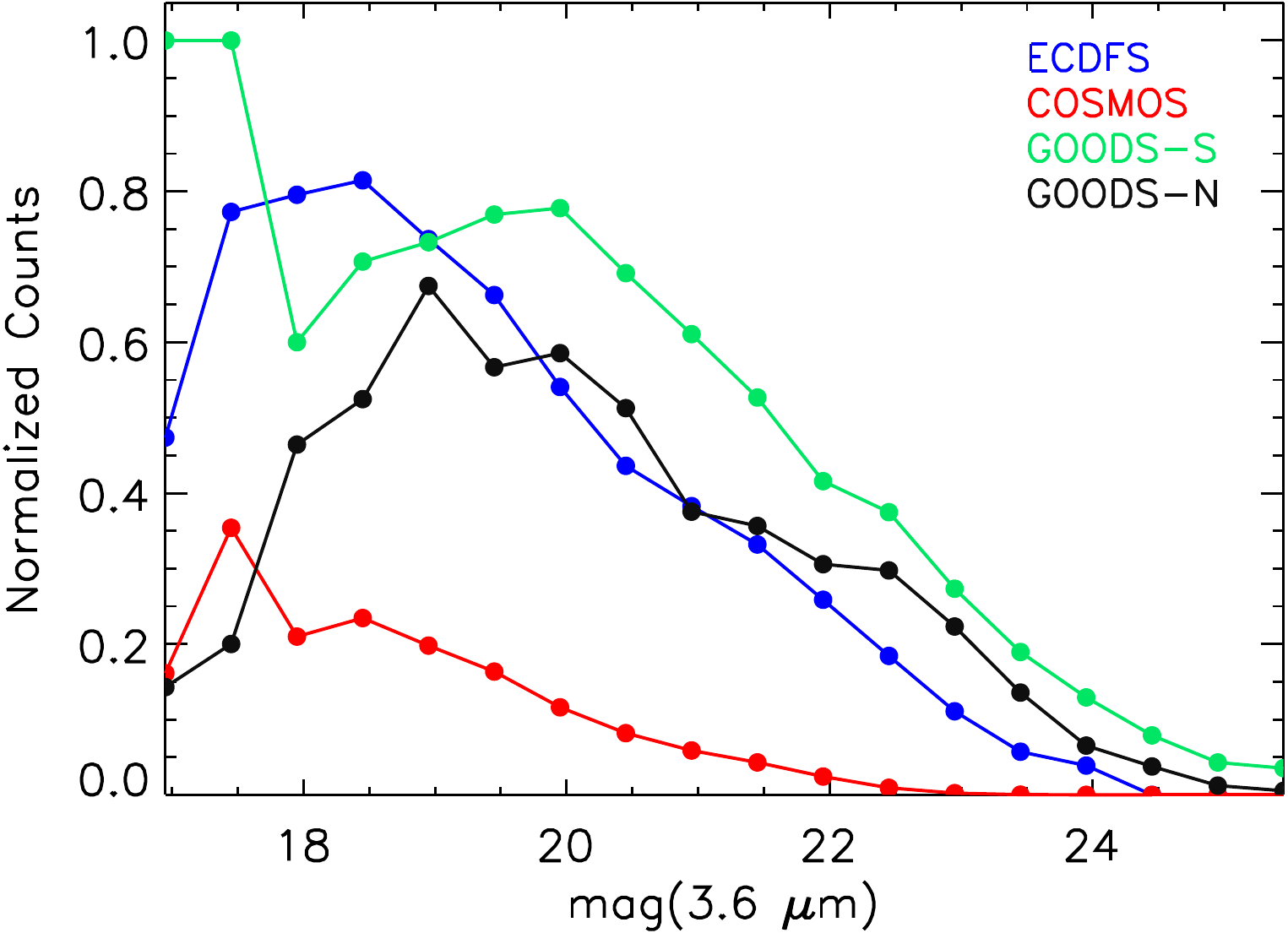}
\includegraphics[width=0.48\hsize]{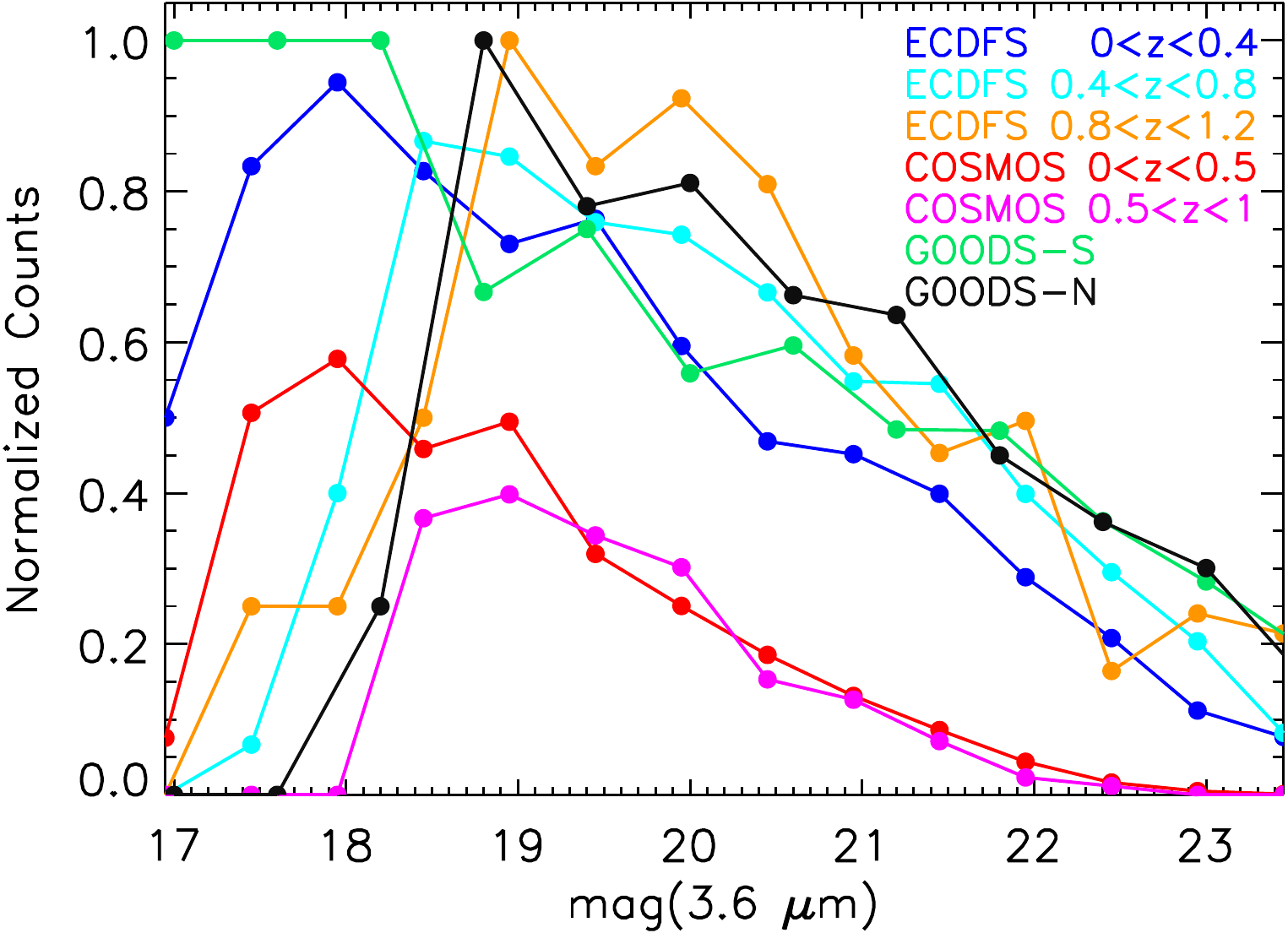}

\caption{Spectroscopic completeness in IRAC 3.6~\um band for the field (on the left) and groups (on the right) in our sample.}
\label{fig:compl_36um}
\end{figure*}

The E-CDFS is one of the best-studied extragalactic fields in the sky (e.g. \citealt{Rixetal2004}; \citealt{Lehmeretal2005}; \citealt{Quadrietal2007}; \citealt{Milleretal2008}; \citealt{Padovani2009}; \citealt{Cardamoneetal2010}; \citealt{Xue2011}; \citealt{Damenetal2011}) with observations from X-ray to radio wavelengths. The smaller Chandra Deep Field South (CDFS, $\alpha = 03h 32m25s$ , $\delta = −27^o 49 ^m 58^s$), in the central part of ECDFS, is currently the deepest X-ray survey with $Chandra$ (4Ms; \citealt{Xue2011}) and XMM-$Newton$ (3Ms; \citealt{Comastri2011}) programmes. In addition to the deep multi-wavelength photometric coverage, the ECDFS has been targeted by many deep spectroscopic surveys. Recently, \cite{Cardamoneetal2010} and \cite{Cooper2011} provide a compilation of all existing high quality redshifts and new IMACS (Inamori-Magellan Areal Camera \& Spectrograph) spectroscopic redshifts in the ECDFS and CDFS, respectively. They reach a spectroscopic completeness down to 
$R \sim 24$~mag similar to that of smaller deep fields such as the GOODS fields \citep{Barger2008} and much higher than larger blank fields such as COSMOS \citep{Lillyetal2007,Lilly2009}, as shown in Fig.~\ref{fig:compl_36um}.

We take the multi-wavelength photometric data from the catalogue of  \cite{Cardamoneetal2010}, which combines a total of 10 ground-based 
broad-band imaging ($U$, $U38$, $B$, $V$, $R$, $I$, $z$, $J$, $H$, $K$), four IRAC imaging ($3.6~\mu$m, $4.5~\mu$m, $5.8~\mu$m, $8.0~\mu$m), and 18 medium-band imaging ($IA427$, $IA445$, $IA464$, $IA484$, $IA505$, $IA527$, $IA550$, $IA574$, $IA598$, $IA624$, $IA651$, $IA679$, $IA709$, $IA738$, $IA767$, $IA797$, $IA856$) filters. The catalogue provides multi-wavelength SEDs and photometric redshifts for $\sim 80000$ galaxies down to $\rm{R_{[AB]}} \sim 27$.  

The spectroscopic galaxy catalogue used for the group member identification in ECDFS is created by combining all available high-quality spectroscopic redshifts. In particular, we use the spectroscopic compilation provided in \cite{Cardamoneetal2010} and more recent spectroscopic catalogues such as \cite{Silverman2010} and the Arizona CDFS Environment Survey (ACES; \citealt{Cooper2011}). \cite{Silverman2010} carried out a program to acquire high-quality optical spectra of X-ray sources detected in the ECDFS up to $\rm{z} = 4$. They measure redshifts for 283 counterparts to Chandra sources using multi-slit facilities on both the VLT (VIMOS, using the low-resolution blue grism with a resolution $\rm{R = 180}$)  and Keck (Deep Imaging Multi-object Spectrograph, DEIMOS; \citealt{Faber2003}). The ACES \citep{Cooper2011} is a recently completed spectroscopic redshift survey of the CDFS conducted using IMACS on the Magellan-Baade telescope. The total number of secure redshifts in the sample is 5080 out of 7277 total, unique targets. The ACES catalogue has a high number of repeated observations. These provide an accurate estimate of the precision of redshift measurements, which have a scatter of  $\sigma \sim$ 75 $\rm{km}~\rm{s}^{-1}$ within the ACES sample \citep{Cooper2011}.  

We remove redshift duplications by matching the \cite{Cardamoneetal2010} catalogue with the \cite{Cooper2011} and the \cite{Silverman2010} catalogues within $1\arcsec$ and by keeping the most accurate ${\rm z_{spec}}$ (smaller error and/or higher quality flag) in case of multiple entries.  Our new ECDFS ${\rm z_{spec}}$ catalogue comprises 7246 unique spectroscopic redshifts.  The 
compilation is culled of candidate stars according to the flags provided in the \cite{Cardamoneetal2010} catalogue: the SExtractor parameter \citep{BertinArnouts1996} enables the selection of non-stellar sources  (we choose $class\_ star<0.95$) and the star flag indicates all the sources for which the best-fitting template is the SED of a star \citep{Cardamoneetal2010}.  The resulting spectroscopic completeness as a function of the IRAC band magnitude at 3.6~$\mu$m is shown in Fig.~\ref{fig:compl_36um} (blue curve for the ECDFS). The completeness is extremely high (80\%) down to $\sim 18.5\ $mag and it is higher than 50\% up to 20 mag.

\subsection{The GOODS fields}

The GOODS data set  \citep{Giavalisco2004} covers approximately 300 arcmin$^2$ divided into two fields: the Hubble Deep Field-North (HDFN) and the CDFS. These fields are the sites of the deepest observations from Hubble, Chandra,  XMM-$Newton$, and many ground-based facilities. GOODS incorporates a Spitzer Legacy project with imaging at 3.6-8~$\mu$m with IRAC and deep 24 $\mu$m imaging with MIPS.

The GOODS-S  field is an area of $15 \times 16$ ${\rm{arcmin}}^2$ within the ECDFS. It has been deeply observed in the X-ray in the 4Ms Chandra and 3 Ms XMM-$Newton$ observations of the CDFS. It has also been targeted by a deep imaging campaign in the optical and near-infrared with the ESO (European Southern Observatory) telescopes \citep{grazian06}. In this work we use the version of the MUSIC (MUltiwavelength Southern Infrared Catalog) catalogue released by \cite{grazian06} to avoid stars, which are properly flagged. The \cite{grazian06}  catalogue is then matched to our own spectroscopic master catalogue of the ECDFS with the addition of GMASS (Galaxy Mass Assembly ultra-deep Spectroscopic Survey; \citealt{Cimatti2008}) redshifts to identify all the members of the \cite{Kurk2009} $z=1.6$ structure.

The GOODS-N field has roughly the same area as the southern counterpart. We use the multi-wavelength catalogue of GOODS-N built by the PEP team \citep{Berta2010} who adopted the \citet{grazian06} approach for the PSF matching. The catalogue includes ACS $bviz$ 
\citep{Giavalisco2004}, Flamingos $JHK$, and Spitzer IRAC data. Moreover, deep $U$, $Ks$ \citep{Barger2008} and MIPS 24~$\mu$m \citep{magnelli09} imaging, and spectroscopic redshifts have been added.


\subsection{The COSMOS survey}
COSMOS is centred on an area of the sky where Galactic extinction is low and uniform ($<$20\% variation; \citealt{Sanders2007}). 
This survey has broad spectral coverage and the imaging survey is complemented by many spectroscopic programs at different telescopes. 
The spectroscopic follow up is still ongoing and so far it includes: Magellan/IMACS \citep{Trump2007} and MMT \citep{Prescott2006} campaigns, the zCOSMOS survey at VLT/VIMOS \citep{Lillyetal2007,Lilly2009}, observations at Keck/DEIMOS (PIs: Scoville, Capak, Salvato, Sanders, Kartaltepe) and FLWO/FAST \citep[][]{Wright2010}. 

The COSMOS photometric catalogue  includes multi-wavelength photometric information for $\sim 2\times 10^6$ galaxies in the entire field.  
We use the catalogue compiled by \cite{Ilbert2009} and \cite{ilbertetal2010}, who cross-match the S-COSMOS  3.6~\um selected catalogue \citep{Sanders2007} with the rest of the multi-wavelength photometry \citep{Capak2007,Capak2009}. They compute photo-z, stellar masses and SFR for all 3.6~\um selected sources \citep{Ilbert2009, ilbertetal2010}.


\subsection{Infrared data}

All the fields considered in this analysis have been observed with {\it{Spitzer}} MIPS at 24 $\mu$m and with {\it{Herschel}} PACS at 100 and 160 $\mu$m. They are all part of the PEP survey, one of the major $Herschel$ Guaranteed Time (GT) extragalactic projects. The ``wedding cake'' structure of this survey, based on four different depths, enables the combination of deep pencil-beam fields with wider, but shallower, areas with better statistics for brighter sources. Indeed, as shown in Table 1, the relatively small GOODS fields reach a much deeper flux detection threshold than the wider ECDFS and COSMOS fields.  In particular, the GOODS fields have also been deeply observed by the GOODS-H survey. This covers a smaller central portion of the entire GOODS-S and GOODS-N regions. Recently the PEP and the GOODS-H teams combined the two sets of PACS observations to obtain the deepest ever available PACS maps \citep{magnelli2013} of both fields. 

For all the fields, we use the PEP source catalogues obtained by applying prior extraction as described in \cite{Lutz2011}. Namely, MIPS 24~$\mu$m source positions are used to detect and extract PACS sources at both 100 and 160~$\mu$m. This is feasible since extremely deep MIPS 24~$\mu$m observations are available for all the fields considered in this work. For each field the source extraction is based on a PSF-fitting technique, presented in detail in \cite{magnelli09}.

In order to take advantage of the much deeper PACS and MIPS observations in GOODS-S with respect to the ECDFS, we use PEP and GOODS-H data in  the GOODS-S field area and the PEP-ECDFS catalogue in the remaining area.

\begin{table}
\centering
\begin{tabular}{l c c c c c}
\hline
Field    & Band 	& Eff. area		& 3$\sigma$ \\
		& 	&			& (mJy)	\\
\hline
GOODS-N &	100 \um	& 187 arcmin$^2$	& 3.0	\\
GOODS-N &	160 \um	& 187 arcmin$^2$ 	& 5.7	\\
GOODS-S &	70 \um 	& 187 arcmin$^2$	& 1.1	\\
GOODS-S &	100 \um	& 187 arcmin$^2$	& 1.2	\\
GOODS-S &	160 \um	& 187 arcmin$^2$	& 2.4	\\
ECDFS &		100 \um	& 0.25 deg$^2$		& 3.9	\\
ECDFS &		160 \um	& 0.25 deg$^2$		& 7.5	\\
COSMOS &	100 \um	& 2.04 deg$^2$		& 5.0	\\
COSMOS &	160 \um	& 2.04 deg$^2$		& 10.2	\\
\hline
\end{tabular}
\caption{ Main properties of the PEP fields used in this work. The first column shows the name of the blank field, the second column the PACS band in which the field is observed, the third column the area covered, and the fourth column the 3$\sigma$ limit in mJy.
}
\label{tab:PEPfields}
\end{table}


\subsection{X-ray data and group selection}
\label{s:xray_analysis}

All the blank fields considered in our analysis have been observed extensively in the X-ray with $Chandra$ and XMM-$Newton$. The data reduction is performed in a homogeneous way, as presented in \cite{Finoguenov2009} and Finoguenov et al. (in preparation). 

Briefly, point sources were subtracted from $Chandra$ and XMM-$Newton$ data sets separately before co-adding them, to allow for source variability.  
The resulting ``residual'' image, free of point sources, is then used to identify extended emission. When these emitting sources have a significance of at least 4$\sigma$ with respect to the background both the presence of a red sequence and spectroscopic redshifts are used to identify galaxy groups. 

The flux is estimated within the largest possible aperture allowed by the background or source confusion, typically  exceeding half of $\mathrm{R_{200}}$. The source flux is corrected for the flux outside the aperture through the use of the beta-model \citep{Cavaliere_FuscoFermiano1976}, as described in \cite{Finoguenov2007}.
The X-ray luminosity  $\mathrm{L_X}$  is estimated within a distance of $\mathrm{R_{200}}$ from the X-ray centre. The X-ray masses $\mathrm{M_{200}}$ are estimated based on the measured $\mathrm{L_X}$, using the scaling relation of \cite{Leauthaud2010}.
The intrinsic scatter in this relation is 20\% (Finoguenov et al. in preparation) and it is larger than the formal statistical error associated with the measurement of $\mathrm{L_X}$. The $\rm L_X$-T relation of \cite{Leauthaud2010} is then used to estimate the temperature, needed also for the computation of the $k-correction$ of the X-ray flux  \cite[see also][]{Finoguenov2007}.


\subsubsection{The group sample}
\label{group_sample}
The main sample of galaxy groups is taken from \cite{popesso2012}. This catalogue comprises 28 groups in the COSMOS field, all at $z < 0.5$ (with the exception of two systems at $z \sim 0.7-0.8$), and 2 groups in the GOODS-N field at $z=0.85$ and $z=1.05$, respectively. 
In order to increase the number of high-redshift groups, we extend the analysis performed on the COSMOS field by \cite{popesso2012} also to the (E)CDFS region. As in \cite{popesso2012}, we create a clean sample of isolated X-ray groups by discarding those showing more than one peak  of similar strength in the spectroscopic redshift distribution and within $3 \times \mathrm{R_{200}}$ from the X-ray group centre, and rejecting all groups with an obvious  close companion. In the former case the redshift association is doubtful, and in the latter case a close companion can strongly bias the estimate of the velocity dispersion and membership. In addition, we choose all groups with at least 10 members in order to obtain a reliable estimate of the velocity dispersion and, thus, the membership. For more details about the computation of group members see the next section, \cite{popesso2012}, and \cite{Biviano+06}.  

We impose a velocity dispersion cut at $\sigma < 1200$ $\rm{km/s}$ to define a 
clear group catalogue and to avoid contamination by massive clusters, whose galaxy population could follow a different evolutionary path \citep{popesso2012}.  Our selection criteria lead to a final number of 22 groups in the ECDFS out of 50 purely X-ray selected groups. 

We consider also a ``super-group'' or large-scale structure spectroscopically confirmed at $z\sim 1.6$ by  \cite{Kurk2009} and dynamically studied by \cite{popesso2012}. 
Due to the different properties of this structure with respect to the group sample considered in this work, we discuss our results in a dedicated section (Sec.~\ref{kurk_sec}).

\begin{figure*}
\centering
   \includegraphics[width=0.48\hsize]{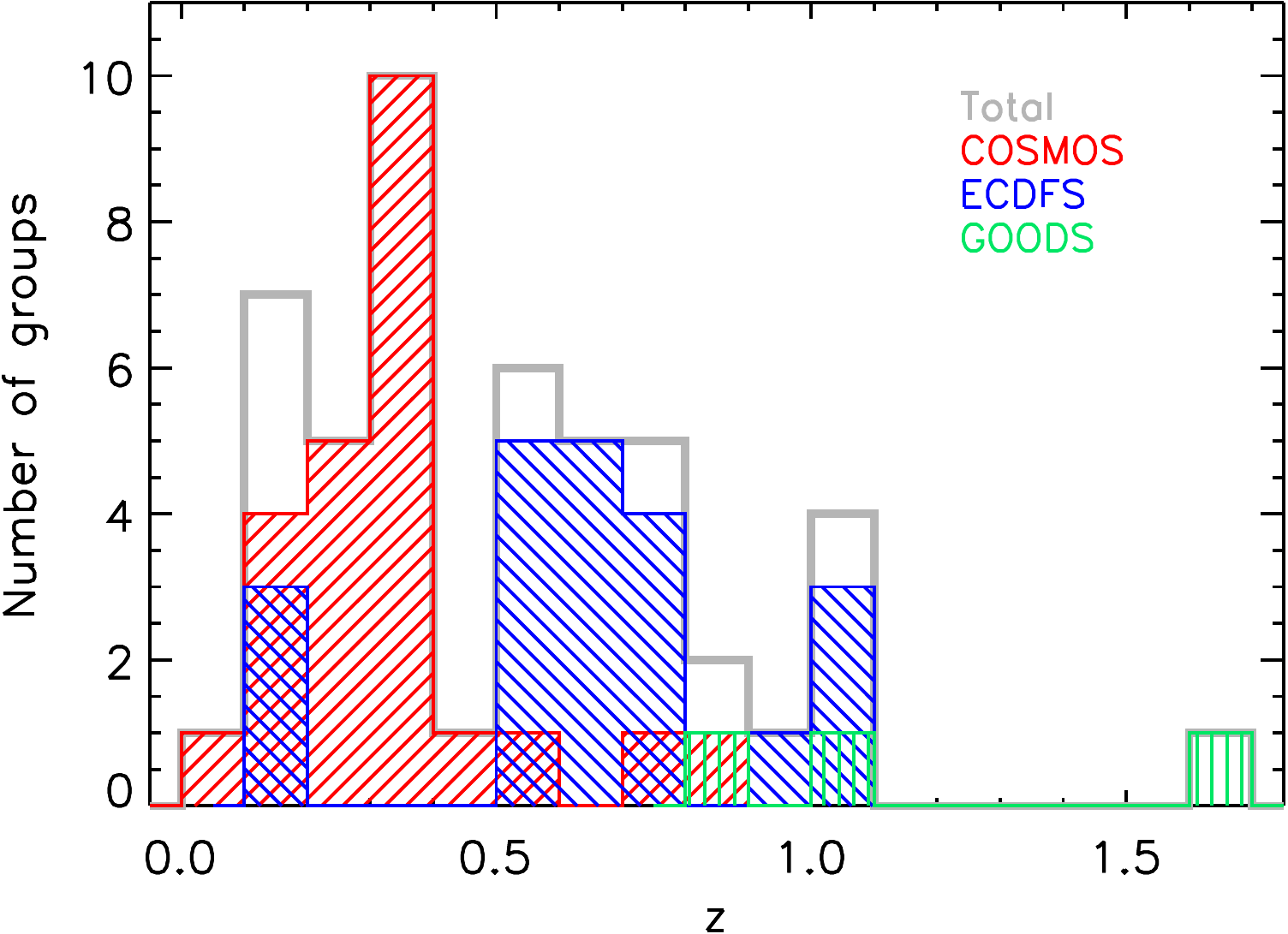}
   \includegraphics[width=0.48\hsize]{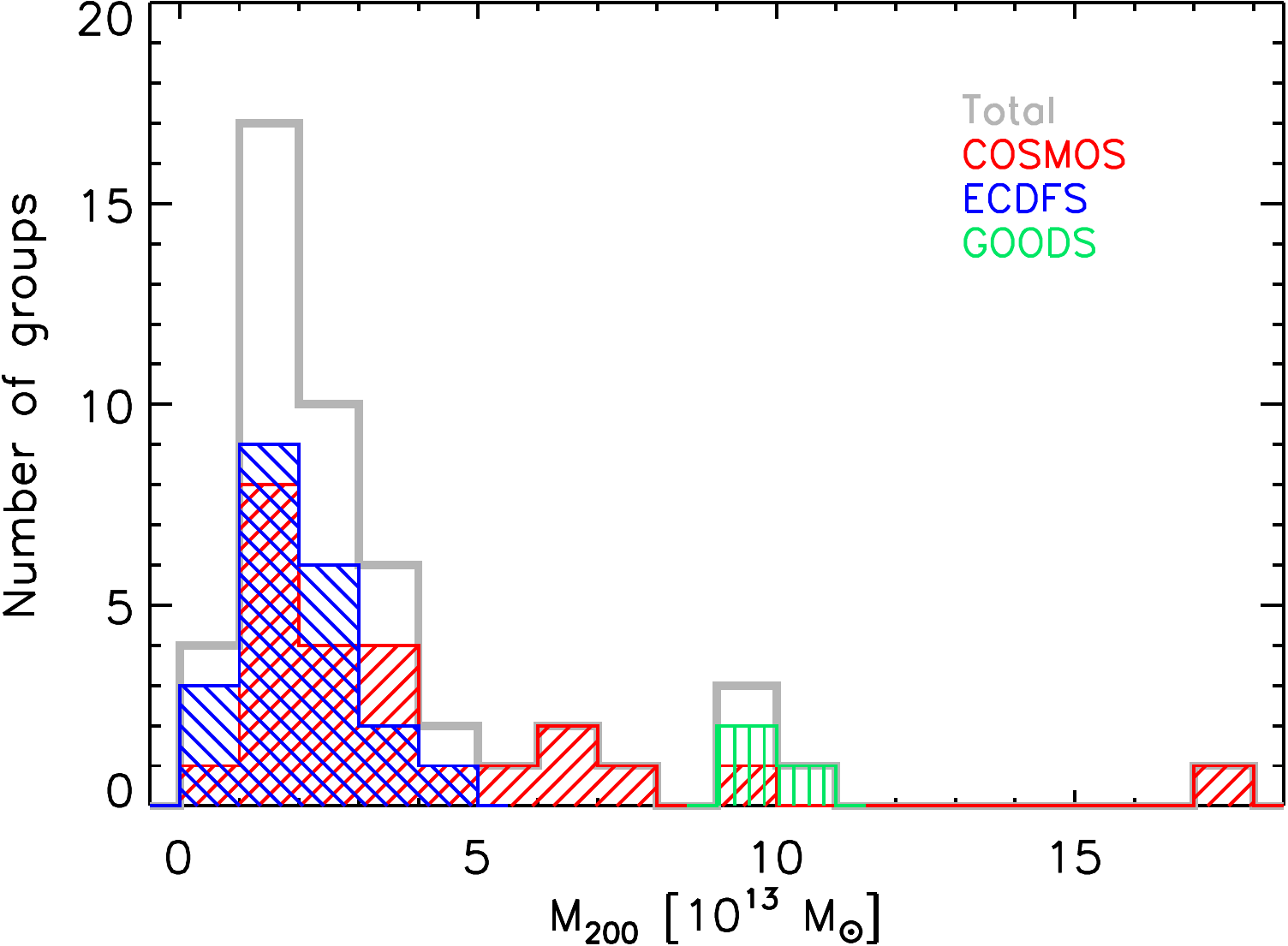}
  \caption{Redshift (left-hand panel) and mass (right-hand panel) distribution of our group sample. Different colours represent the different surveys to which the groups belong.}
  \label{z_histo}
\end{figure*}

The redshift and X-ray mass distribution of the sample is shown in Fig. \ref{z_histo}. The mass distribution peaks at $\rm 2\times 10^{13}~M_{\odot}$. We checked that every redshift bin is populated by groups with similar mean total mass. The last redshift bin hosts the $z \sim 1.6$ super-structure which we use for comparing our results at higher redshift.


\section{Membership and galaxy properties}
\label{calibration}
This section describes how galaxies are classified as group members via dynamical analysis for each extended source of X-ray emission. We also show how the SFRs and stellar masses are estimated for each group galaxy member. Our final aim is to perform the analysis of the evolution of the SF activity in the group environment.

\subsection{Membership} 
\label{s:members}
The galaxy membership is based on the {\it Clean} algorithm of
\citet{Mamon+12} which is based on modelling of the mass and
anisotropy profiles of cluster-sized haloes extracted from a
cosmological numerical simulation. After selecting the main group
peak in redshift space by the method of weighted gaps, the algorithm
estimates the group velocity dispersion using the galaxies in the
selected peak. This is then used to evaluate the virial velocity based
on assumed models for the mass and velocity anisotropy profiles. These
models with the estimated virial velocity are then used to predict the
line-of-sight velocity dispersion of the system as a function of
system-centric radius, $\sigma_{\rm{los}}$(R).  Any galaxy having a
rest-frame velocity within $\pm 2.7 \sigma_{\rm{los}}$(R) at its
system-centric radial distance R is selected as group member.  We
use the X-ray surface-brightness peaks as centres of the X-ray
detected systems.  The new members are used to re-compute the global
group velocity dispersion, hence its virial velocity, and the
procedure is iterated until convergence.  The value of the virial
velocity obtained at the last iteration of the {\it Clean} algorithm is used
to evaluate the system dynamical mass.

In \cite{popesso2012} the dynamical and X-ray mass
estimates are in good agreement in the COSMOS field. 
We note much less agreement for the newly defined (E)CDFS group sample, where the dynamical masses are on average higher than the X-ray masses. This could be due to the fact that ECDFS groups are on average much more distant than COSMOS groups, and this is only partially explained by the deeper X-ray exposure in the ECDFS field with respect to the COSMOS field. In the following we use the X-ray masses for all systems for which they are available, since unlike dynamical masses they do not suffer from projection effects, which may be considerable when the number of spectroscopic members is low, as in our sample.

\subsection{Infrared luminosities}
\label{computation_ir_lum_and_sfr}

We compute the IR luminosities ($\rm L_{IR}$) by fitting the photometry with the recent SED templates presented by \cite{Elbaz2011} and integrating them over the range 8-1000\,$\mu$m. The PACS (100 and 160\,$\mu$m) fluxes, when available, together with the 24~\um fluxes are used to find the best-fitting templates among the main sequence (MS) and starburst (SB; \citealt{Elbaz2011}) templates. When only the 24~\um flux is available for undetected PACS sources, we rely only on this single point and we use the MS template for extrapolating the $\rm L_{IR}$. Indeed, the MS template turns out to be the best-fitting template in the majority of the cases (80\%) with both PACS and 24~\um detection. 

\begin{figure}
  \centering
  \includegraphics[width=\hsize]{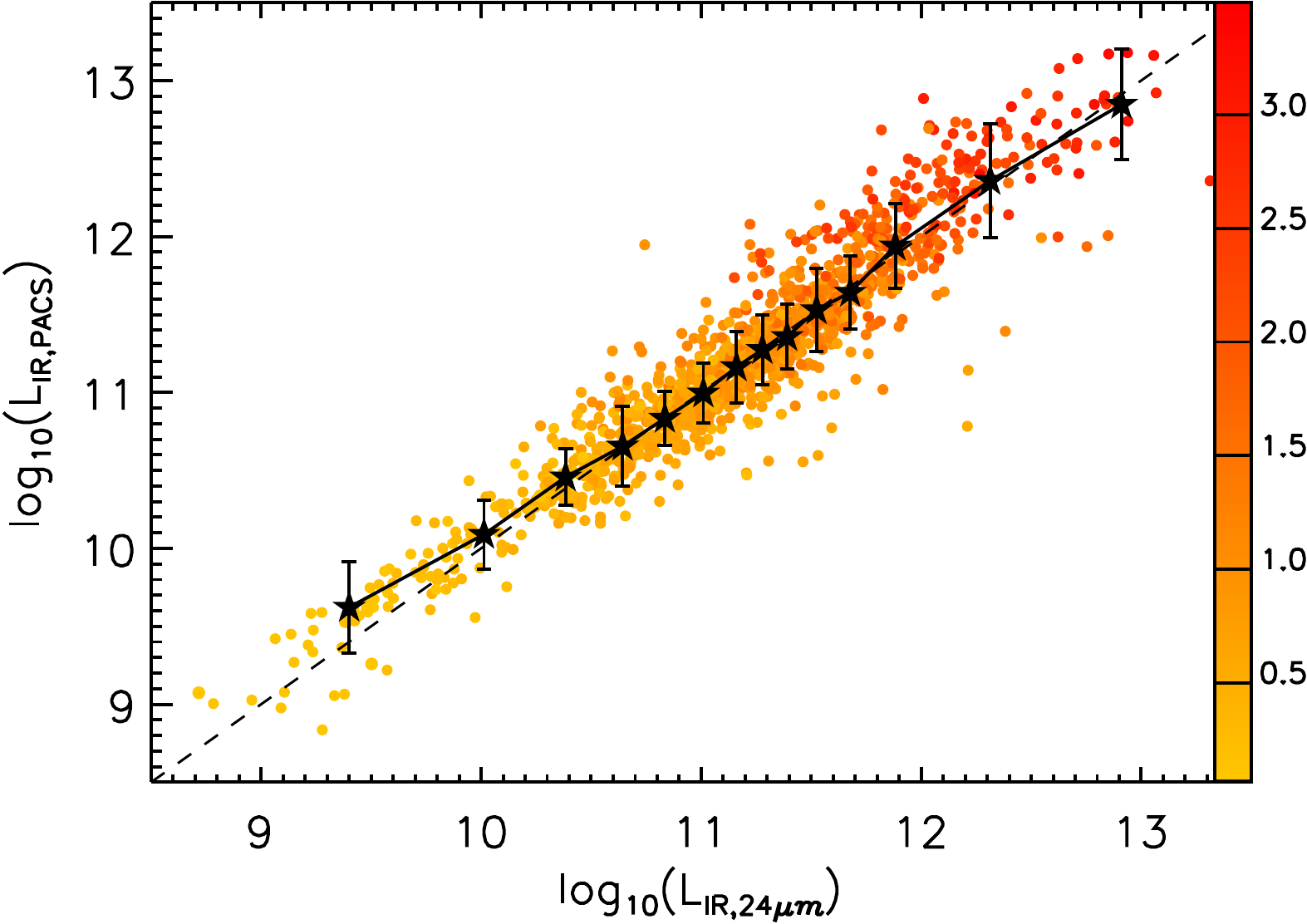}
  \caption{Comparison between the extrapolation of $\rm L_{\rm IR}^{\rm tot}$ from PACS (100 and 160\,$\mu$m) versus that from 24\,$\mu$m for the GOODS fields as a function of redshift (colour bar).  The dashed line represents a one to one relation while the black stars with error bars represent the median infrared luminosity based on $Herschel$ plus 24~\um fluxes for each bin of $\rm L_{\rm IR}^{\rm 24}$. The luminosities are expressed in solar units. 
}
  \label{ir_comp_goods}
\end{figure}

In principle, the use of the MS template could cause only an underestimate of the extrapolated $\rm L_{IR}$ from 24~\um fluxes, in particular at high redshift or for off-sequence sources ($\rm L_{IR}^{24} > 10^{11.7}\ L_\odot$). This is due to the relatively higher PAHs emission of the MS template (see \citealt{Elbaz2011} for more details). 
However, the $\rm L_{IR}$ estimated with the best-fitting templates based on PACS and 24~\um data agrees very well with the $\rm L_{IR}$ extrapolated from the 24~\um flux ($\rm L_{IR}^{24}$) using the MS template.
Fig.~\ref{ir_comp_goods} shows such a comparison for the GOODS fields where we have the deepest PACS coverage. In larger fields such as COSMOS and ECDFS there is a larger probability to find rare strong star-forming off-sequence galaxies even at low redshift. However those sources should be detected by the $Herschel$ observations given their very high luminosity, above the PACS detection threshold. Thus, even for these rare cases we correctly estimate the $\rm L_{IR}$.
Our estimated luminosities are also consistent with those computed using an alternative set of templates from \cite{rodighiero10}.

In our analysis we use the \cite{Kennicutt1998} relation to convert the bolometric infrared luminosity into SFR.  This formula assumes a \cite{Salperter1955} Initial Mass Function (IMF). We apply an offset of -0.18~dex to convert the obtained SFR for the \cite{Chabrier2003} IMF, which is the IMF adopted by our SED fitting procedure.


\subsection{Stellar masses and star formation rates from SED fitting}
\label{sec:SFR_from_SED_fitting}
Due to the flux limit of the MIPS and PACS observations, the mid- and far-infrared data allow us to probe the region of the normally and highly star-forming galaxies that would lie on or above the SFR-mass MS (e.g. \citealt{Noeske2007, Elbaz2007}). To cover also the region below the MS in the SFR-mass diagram we also estimated the SFR and stellar masses from SED fits to the shorter wavelength data.  In this section we describe our procedure to compute these properties for ECDFS and GOODS-South. We use the values computed by \cite{ilbertetal2010} for COSMOS and those of \cite{Wuyts2011} for GOODS-North. 

We compute SFR and stellar masses for ECDFS and GOODS-South using \lephare (PHotometric Analysis for Redshift Estimations; \citealt{arnoutsetal2001, ilbertetal2006}), a publicly available\footnote{http://www.cfht.hawaii.edu/~arnouts/LEPHARE/cfht\_lephare/ lephare.html} code based on a $\mathrm{\chi^2}$ template-fitting procedure.  
We follow the procedure described in \cite{Ilbert2009, ilbertetal2010} 
First we adjust the photometric zero-points, as explained in \cite{ilbertetal2006}.  Namely, using a $\chi^2$ minimization at fixed redshift, we determine for each galaxy the corresponding best-fitting COSMOS templates (included in the package; see \citealt{ilbertetal2006}).  Dust extinction is applied to the templates using the \cite{Calzetti2000} law [with $E(B-V)$ in the range 0 - 0.5 and with a step of 0.1].

We apply the systematic zero-point offsets to our catalogues (ECDFS and GOODS-MUSIC) and compute the SFR and stellar masses using {\it Le PHARE}, following the recipe of \cite{ilbertetal2010}. The SED templates for the computation of mass and SFR are generated with the stellar population synthesis package developed by \citet[BC03]{bruzual_charlot2003}. We assume a universal IMF from \cite{Chabrier2003} and an exponentially declining SF history $SFR \propto e^{-t/\tau}$ (with $\rm 0.1~Gyr <\tau<30~Gyr$). The SEDs are generated for a grid of 51 ages (spanning a  range from 0.1 Gyr to 14.5 Gyr). Dust extinction is applied to the SB templates using the \cite{Calzetti2000} law (with $E(B-V)$ in the range 0 - 0.5 and with a step of 0.1).  Depending on the template we also include emission lines as described in \cite{ilbertetal2010}.

To check the robustness of our estimates we compare them with those computed by Wuyts et al. (in preparation) for the same field and using the method of \cite{Wuyts2011}. Namely, SFR and stellar masses are computed using FAST \citep{Krieketal2009}, a software which searches for the best fit among different templates (and a multi-dimensional grid with different ages, extinctions, $\tau$).  \cite{Wuyts2011} use the BC03 library assuming an IMF from \cite{Chabrier2003} and the same declining SF history we use (restricted to a minimum  $\tau$
of 300 Myr), with a dust extinction from the \cite{Calzetti2000} law.  
Since Wuyts et al.  do not add emission lines to the templates, for the sake of comparison we generate a further set of stellar masses and SFR using our own methods without adding the emission lines. The comparison of the stellar masses provides very good agreement, with a fraction of outliers beyond 3$\sigma$ (5$\sigma$) of 2\% (0.5\%) and a scatter ${\rm \sigma(log\, M)=0.34}$.  The comparison of the SFRs has a somewhat higher scatter of  ${\rm log\, SFR=0.61}$. Similar results are obtained if we compare our estimates with those of \cite{Santini2009} in GOODS-S. 

 \begin{figure*}
  \centering
  \includegraphics[width=0.48\hsize]{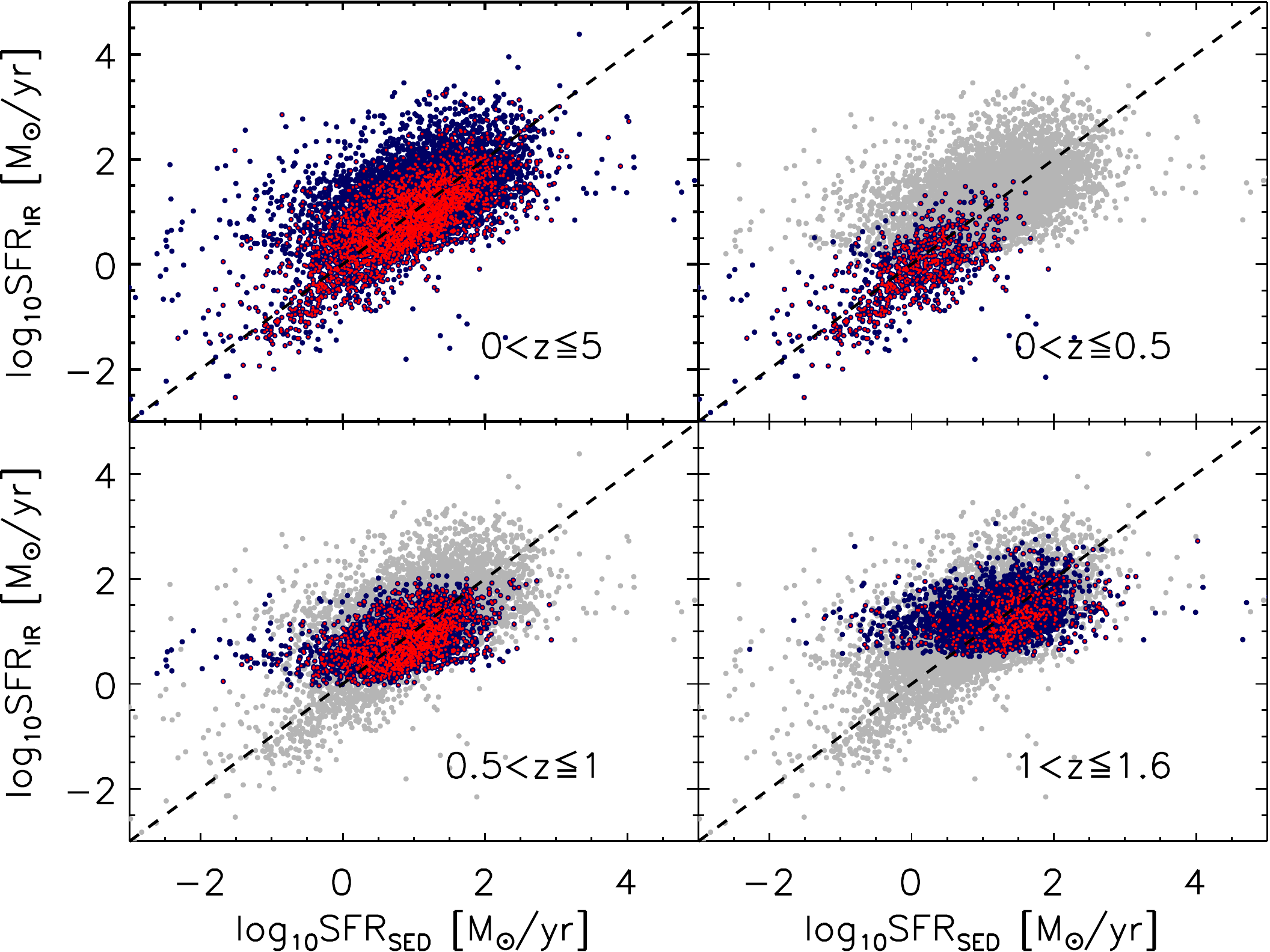}
  \includegraphics[width=0.48\hsize]{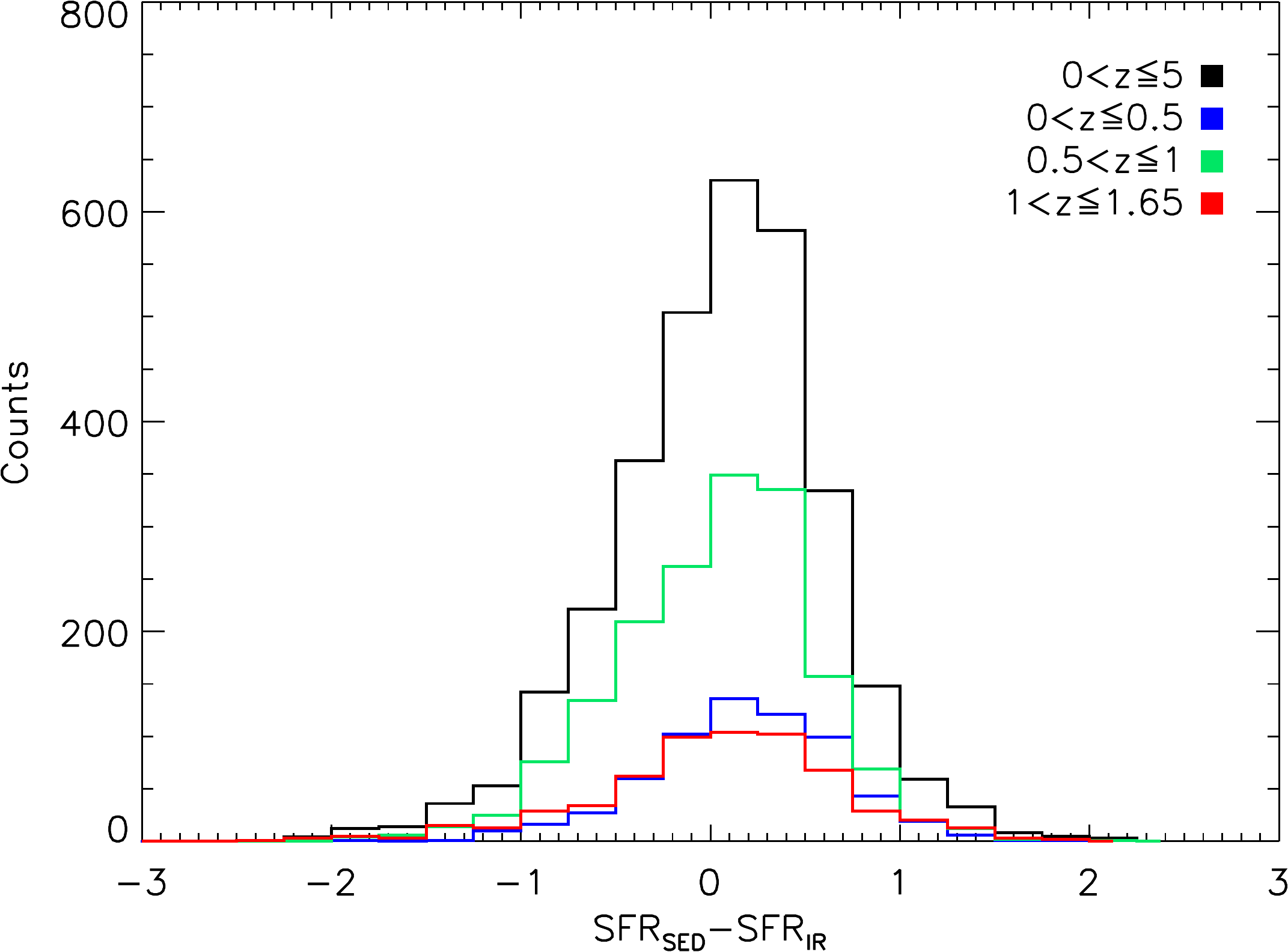}
  \caption{{\it Left}: comparison between $\rm SFR_{IR}$ and $\rm SFR_{SED}$ for the ECDFS and GOODS-S. The upper left panel shows all the sources with an IR detection in the field. The same distribution is represented as grey dots in the following panels, where we show all the sources for different redshift bin with blue dots : ${\rm 0<z\leq 0.5}$, ${\rm 0.5<z\leq 1}$ and ${\rm 1<z\leq 1.6}$, respectively from left to right and top to bottom. In all panels the red dots represent the sources with spectroscopic redshifts, while the dashed line is the one-to-one relation. {\it Right}: histograms of $\rm SFR_{SED}-SFR_{IR}$ residuals for all the galaxies with spectroscopic redshift. The different colours correspond to the same redshift bins used in the left-hand panel. All the histograms peak around 0. We measure a scatter of 0.61~dex for the whole range of redshifts, 0.58~dex for ${\rm 0<z\leq 0.5}$, 0.57~dex  for ${\rm 0.5<z\leq 1}$ and 0.67~dex  for ${\rm 1<z\leq 1.6}$ for all sources with spectroscopic redshift.}
  \label{calib_sfr}
\end{figure*}

As a further check, we calibrate our optical/UV SED-based SFR estimates versus the more robust SFR based on IR emission for the sample of MIPS and/or PACS detected galaxies. Our calibration is done in 3 different redshift bins: ${\rm 0 < z \leq 0.5}$, ${\rm 0.5 < z \leq 1}$, and ${\rm 1 < z \leq 1.6}$. The left-hand panel of Fig.~\ref{calib_sfr} shows the result of this comparison. 
The estimates are broadly consistent, although the scatter is quite large: 0.73~dex for the whole range of redshifts, 0.74~dex for ${\rm 0<z\leq 0.5}$, 0.63~dex for ${\rm 0.5<z\leq 1}$ and 0.68~dex for ${\rm 1<z\leq 1.6}$ for the ${\rm SFR_{SED}-SFR_{IR}}$ relation. The situation improves when only the spectroscopic sample is considered, probably because galaxies with a spectroscopic identification are on average brighter and bluer than the overall sample.  For the spectroscopic sample the scatter (see right-hand panel of Fig.~\ref{calib_sfr}) is 0.61~dex for the whole range of redshifts, 0.58~dex for ${\rm 0<z\leq 0.5}$, 0.57~dex  for ${\rm 0.5<z\leq 1}$ and 0.67~dex  for ${\rm 1<z\leq 1.6}$.

We conclude that our estimates of the stellar mass are accurate within a factor of 2, while the SFR are more difficult to constrain via SED fitting. Indeed, previous studies \citep{Papovich2001,Shapley2001,Shapley2005,Santini2009} already demonstrate that, while stellar masses are well determined, the SED fitting procedure does not strongly constrain SFR and histories at high redshifts, where the uncertainties become larger due to the SFR--age--metallicity degeneracies. In any event, we stress that the SFR derived via SED fitting are used only for MIPS and PACS undetected objects, thus, well below the MS in the redshift range considered in our analysis.


\section{Accounting for spectroscopic incompleteness in the galaxy sample}
\label{sec:completeness_mock_cat}
Since the group members are spectroscopically selected, we need to consider how the spectroscopic selection function drives our galaxy selection and, thus, how it can affect our results. Fig.~\ref{fig:compl_36um} shows the spectroscopic completeness as a function of the apparent 3.6 $\mu$m magnitude. The left-hand panel shows the spectroscopic completeness of the full field area for each survey. The right-hand panel shows the mean spectroscopic completeness in the group regions. This is estimated as the mean of the completeness in the cylinder along the line of sight to each group and within $1.5\times R_{200}$ from the group centre. 
We must account for this incompleteness in order to understand what, if any, selection biases might affect our analysis. 
We bin the sample of groups by redshift to distinguish between the high-redshift groups that happen to be mainly in the GOODS-S area with a somewhat higher spectroscopic completeness than the full ECDFS area, and the three low-redshift groups that reside at the edge of the ECDFS area with lower spectroscopic completeness. The spectroscopic completeness of the COSMOS field is much lower with respect to the other fields both in the full area and in the group area. This is mainly due to the difficulty in efficiently covering the full COSMOS area ($2~\rm{deg}^2$) with spectroscopic follow-up. 

In order to estimate the errors involved in our analysis and check for possible biases due to the spectroscopic incompleteness, we design a method to use the mock catalogue of \cite{KW_millennium2007} drawn from the Millennium run \citep{Springel2005} to simulate a catalogue with a spectroscopic selection function similar to the one observed in the fields considered in our analysis. We briefly describe the procedure in the next Section.

\subsection{The Millennium mock catalogues}
\label{sec:completeness_mock_mass_sfr}
The Millennium simulation follows the hierarchical growth of dark matter structures from
redshift $z=127$ to the present \citep{Springel2005}. Out of several mock catalogues created from the Millennium simulation, we choose to use those of \cite{KW_millennium2007} in order to estimate the errors due to the incompleteness of our spectroscopic catalogues.
The simulation assumes the
concordance $\Lambda$CDM cosmology and follows the trajectories of
$2160^3\simeq 1.0078\times 10^{10}$ particles in a periodic box
500 Mpc $h^{-1}$ on a side. 

\cite{KW_millennium2007} make mock observations of the artificial Universe by positioning a virtual observer at
$z\sim 0$ and finding the galaxies which lie on the appropriate backward light-cone. The backward light-cone is defined as the set of all light-like worldlines intersecting the position of the observer at redshift zero. 

We use the mock catalogues from 2 out of 6 such light cones to take into account field to field variation. We select as information from each catalogue the Johnson photometric band magnitudes available ($R_J$, $I_J$ and $K_J$), the redshift, the stellar mass and the SFR of each galaxy with a cut at $I_J < 26$ to limit the data volume to the galaxy population of interest. In order to simulate the spectroscopic completeness observed in the region of our groups, we randomly extract for each mock catalogue a sub-sample of galaxies by following the spectroscopic completeness of reference. Namely, we choose one of the available photometric bands and extract randomly in each magnitude bin a percentage of galaxies consistent with the percentage of systems with spectroscopic redshift in the same magnitude bin of our galaxy sample in the group region.  
We follow this procedure to extract randomly 50 different catalogues from each light-cone. We end up with 100 different (randomly extracted) catalogues that appropriately reproduce the same characteristics of the selection function of our sample. 

To check this we apply the following approach: taking advantage of the very high accuracy of the photometric redshifts of \cite{Cardamoneetal2010} in the ECDFS, we estimate a spectroscopic completeness in physical properties such as stellar mass and SFR of our galaxy sample (in the group region). We assume the photometric redshifts, and the physical properties based on those, to be correct. Then, we divide our sample into four redshift bins following the separation done for the groups and the analysis presented in the next Section. We then estimate the spectroscopic completeness as a function of stellar mass and SFR in each bin. The spectroscopic completeness is estimated as the ratio of the number of the galaxies with spectroscopic redshift 
to the number of all those with $z_{phot}$ in the considered redshift bin and per bin of stellar mass or SFR. 

\begin{figure}
\begin{center}
\includegraphics[width=0.45\hsize]{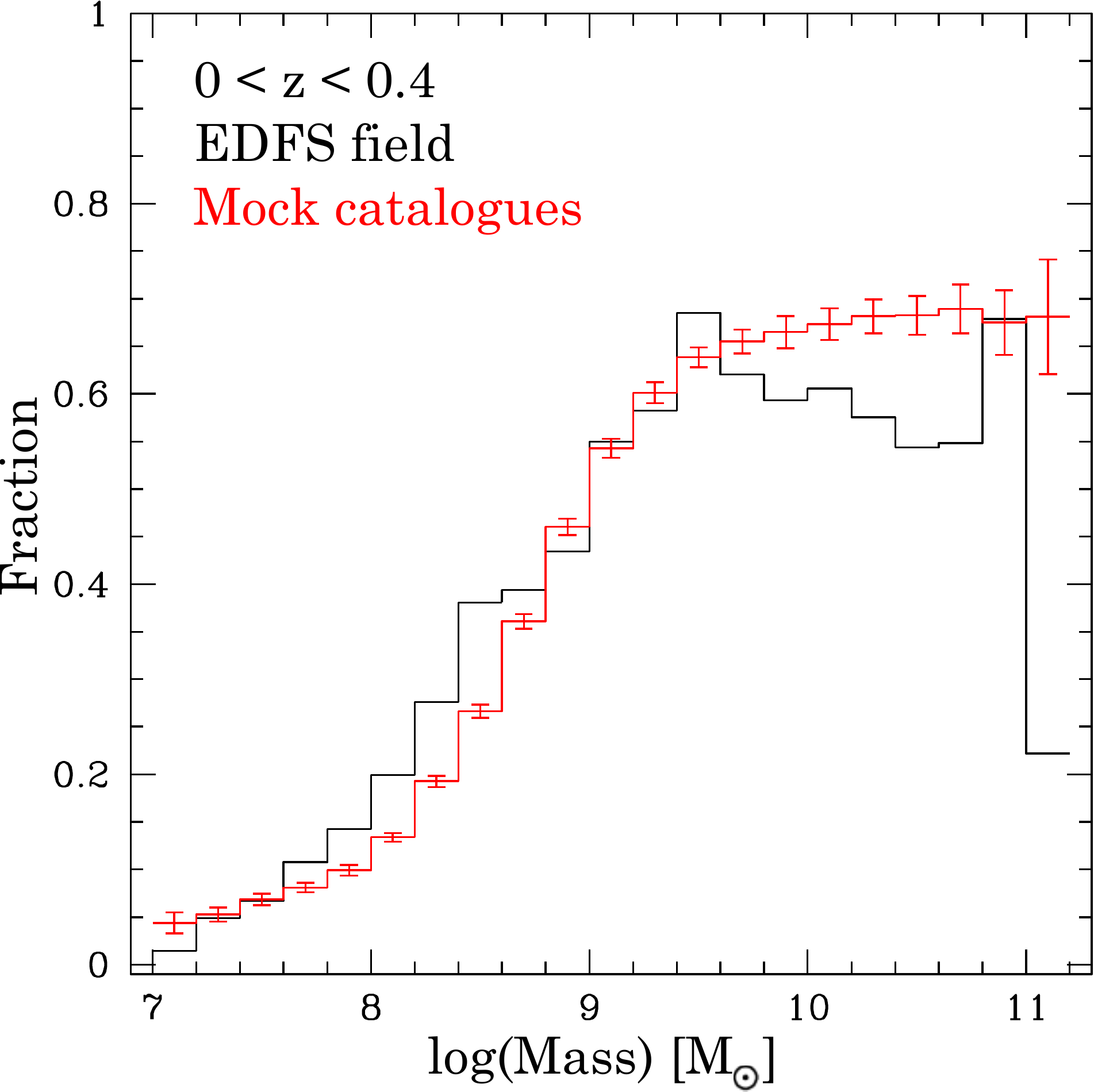}
\includegraphics[width=0.45\hsize]{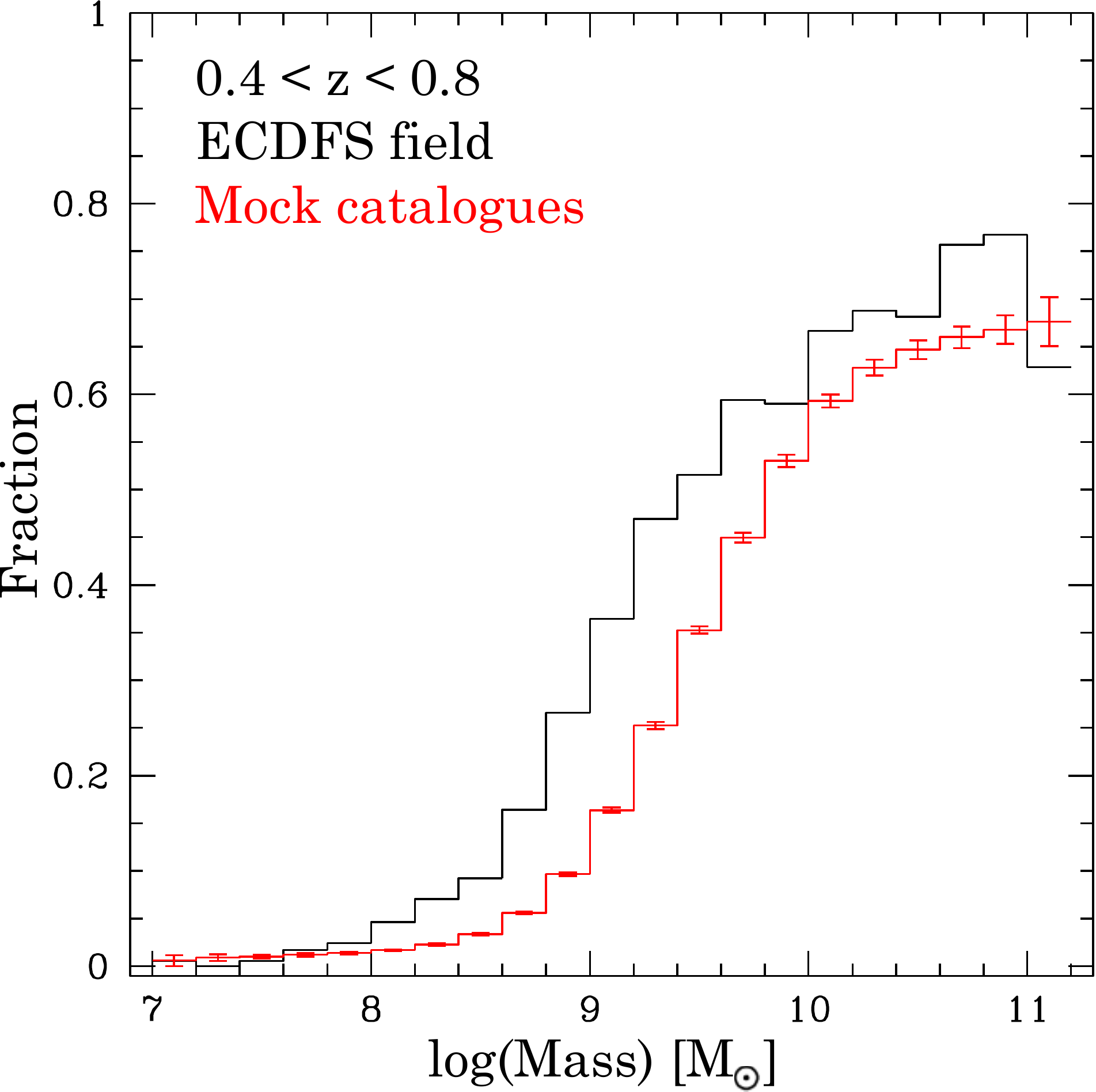}
\includegraphics[width=0.45\hsize]{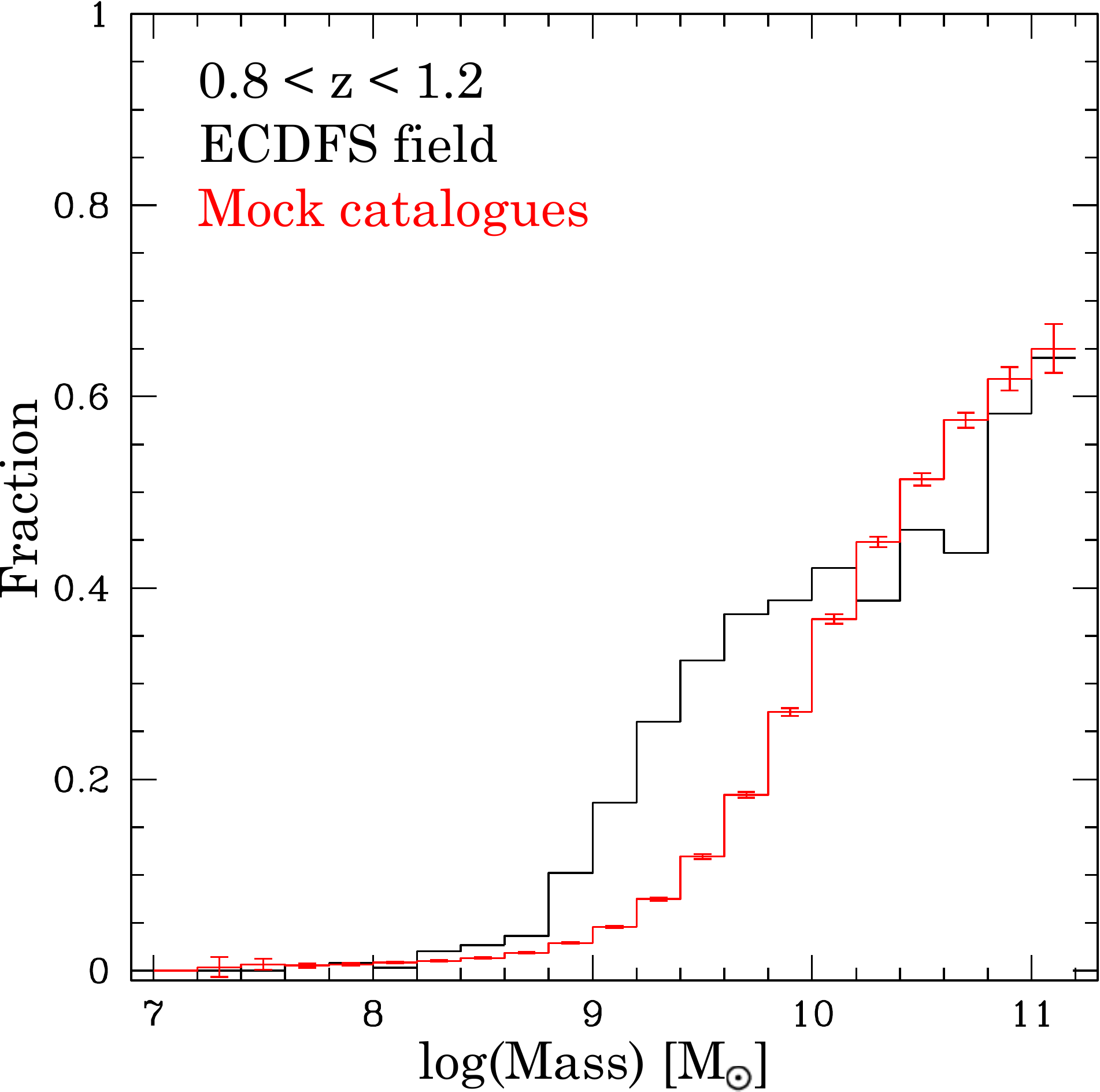}
\includegraphics[width=0.45\hsize]{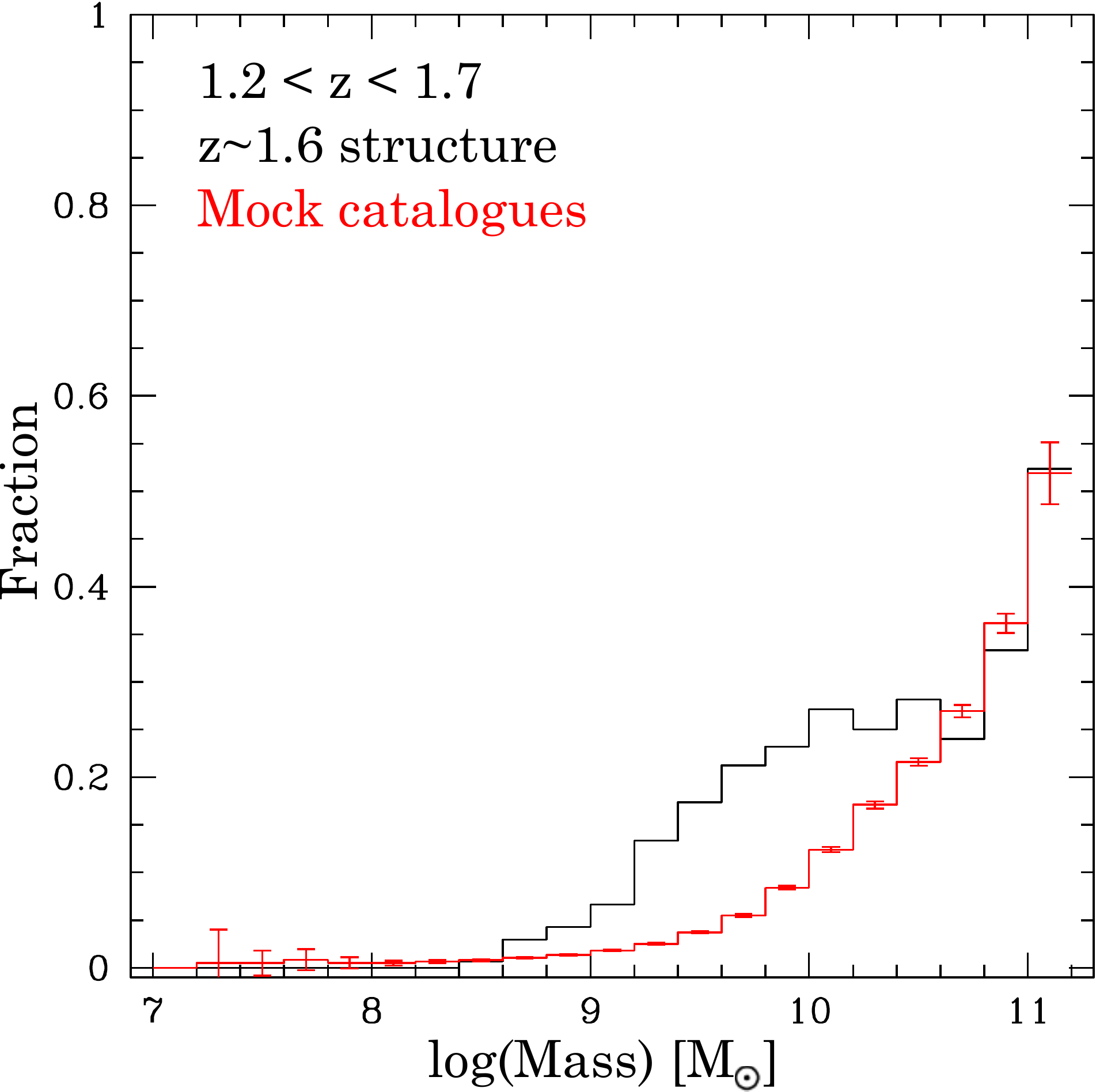}

\end{center}
\caption{Spectroscopic completeness as a function of the galaxy stellar mass for the ECDFS (black histogram) and the mock catalogues (red histogram) in 4 redshift bins.}
\label{mock_compl_mass_ECDFS}
\end{figure}

\begin{figure}
\begin{center}
\includegraphics[width=0.45\hsize]{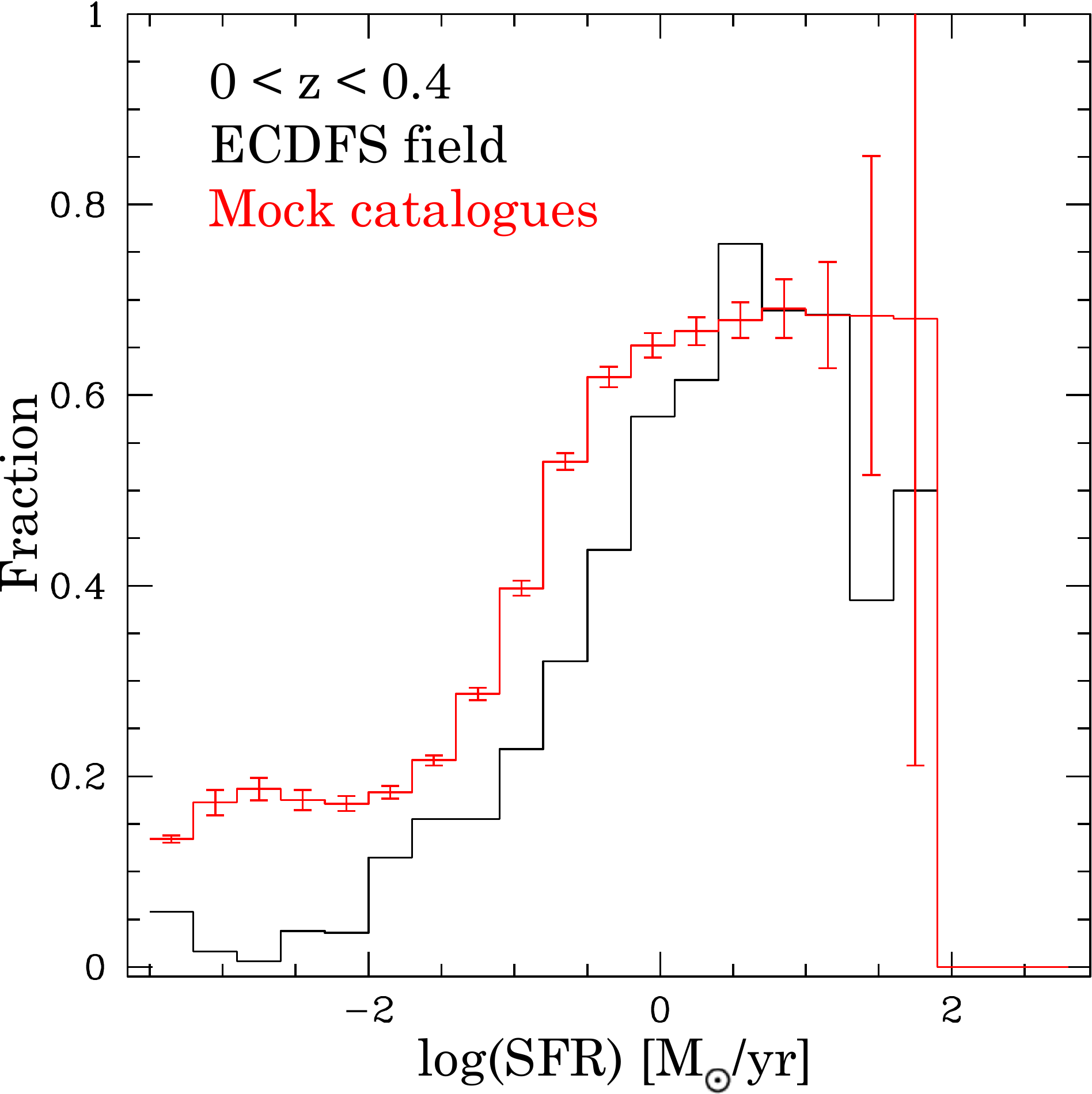}
\includegraphics[width=0.45\hsize]{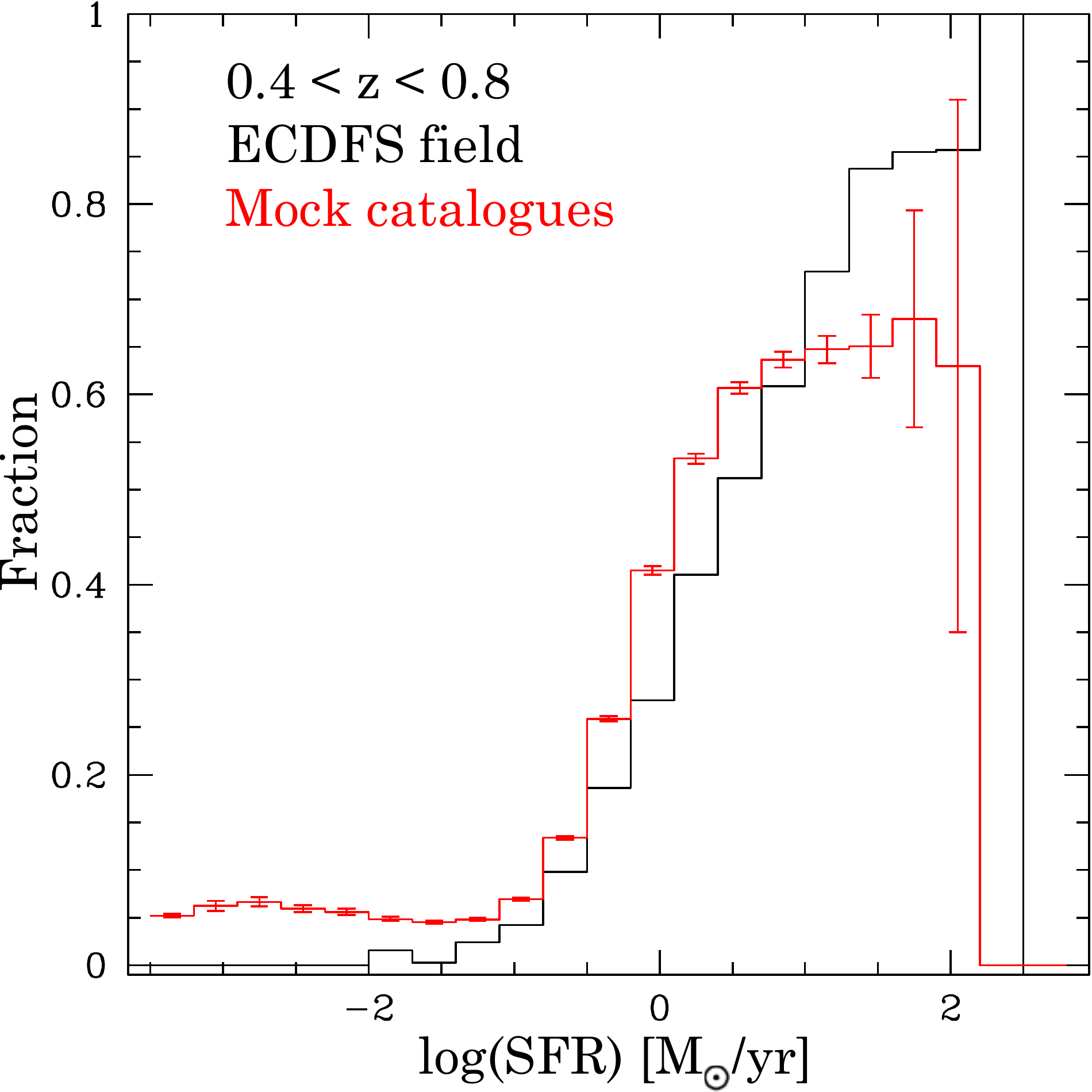}
\includegraphics[width=0.45\hsize]{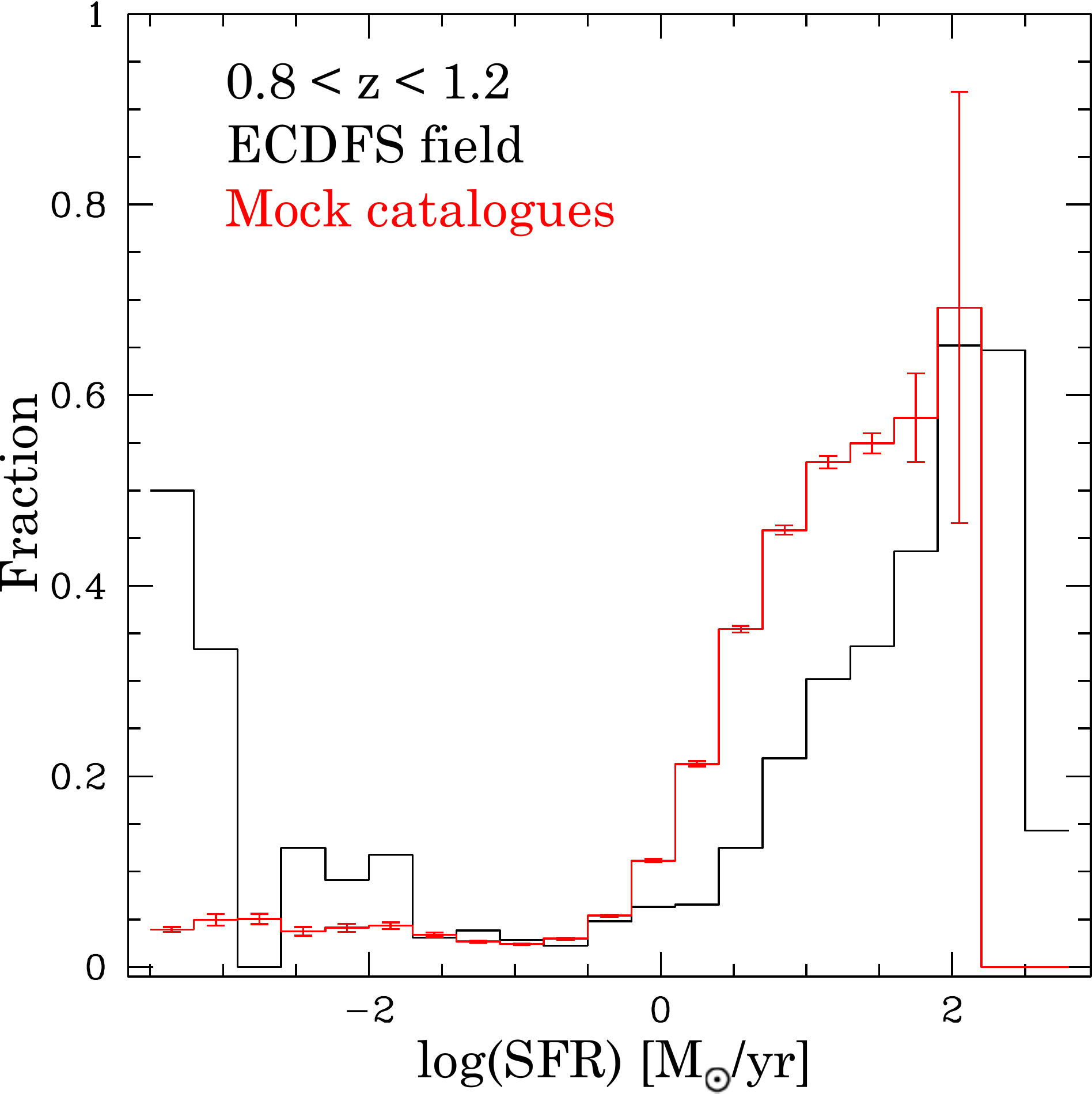}
\includegraphics[width=0.45\hsize]{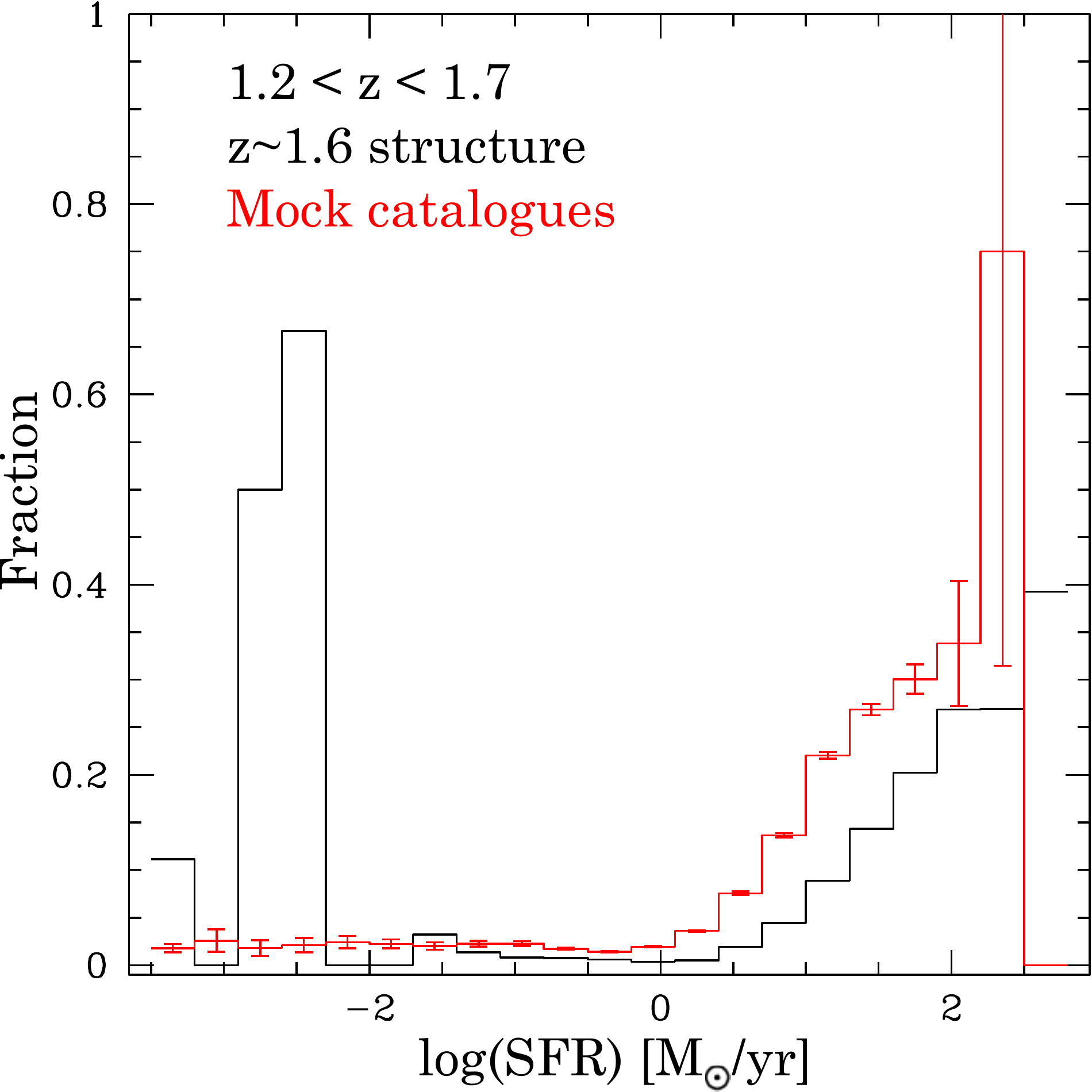}

\end{center}
\caption{Spectroscopic completeness as a function of the galaxy SFR for the ECDFS (black histogram) and the mock catalogues (red histogram) in 4 redshift bins.}
\label{mock_compl_sfr_ECDFS}
\end{figure}

This procedure allows us to determine how the spectroscopic selection, based on the photometric information (e.g. colour, magnitude cuts, etc.), affects the choice of galaxies as spectroscopic targets according to their physical properties. In order to check for possible biases, we follow the same approach for the randomly-extracted mock catalogues. We apply the same redshift bin separation applied to the real catalogue. Then, we estimate in each bin the completeness as the ratio between the number of galaxies in that bin (and per bin of stellar mass and SFR) with respect to the number of galaxies in the parent sample (the original mock catalogue of the same light-cone). 

The comparison between the observed completeness in ECDFS and in the corresponding mock catalogues is shown in Fig.s~\ref{mock_compl_mass_ECDFS} and  \ref{mock_compl_sfr_ECDFS}. We show the mean completeness averaged over the 100 randomly-extracted 
catalogues created following the spectroscopic completeness of the ECDFS in Johnson $R$ band. In all panels the mock catalogues tend to reproduce, to a level that we consider sufficient for our needs, the selection of massive and highly star-forming galaxies observed in the real ECDFS sample. We note a significant difference only in the region of very low SFR where the completeness is much higher in the ECDFS sample than in the mock catalogues at high redshift. This is due to the fact that massive early-type galaxies have been targeted in dedicated observations (see \citealt{Popesso2009} for more details), especially in the GOODS-S field region and at $ z> 0.5$. 

We point out that, while the galaxy mock catalogues of the Millennium simulation provide a suitable representation of the relatively nearby galaxies, at higher redshift ( $z > 1$) they fail in reproducing the correct distribution of star-forming galaxies in the SFR-stellar mass plane, as already shown by \cite{Elbaz2007}. This is related to the difficulty that the semi-analytical models have in predicting the observed evolution of the galaxy stellar mass function and the cosmic SF history of our Universe \citep{KW_millennium2007, Guo2010}.  Indeed \cite{Elbaz2007} estimate that at $0.8 < z < 1.2$ the galaxy SFR is underestimated, on average, by a factor of two,  at fixed stellar mass, with respect to the observed values. By performing the same exercise with our data set, we find that this underestimation ranges between factors of 2.5-3 at $1.2 < z < 1.7$. 

However, we stress here that this does not represent a problem for our approach. Indeed, the aim of using the Millennium galaxy mock catalogues is to understand what is the bias introduced by a selection function similar ``in relative terms'' to the spectroscopic selection function caracterizing our data set. In other words, for our needs it is sufficient that the randomly-extracted mock catalogues reproduce the same bias in selecting, on average, the same percentage of most star-forming and most massive galaxies of the parent sample. The bias of our analysis will be estimated by comparing the results obtained in the biased randomly-extracted mock catalogues and the unbiased parent catalogue.  Since the underestimation of the SFR or the stellar mass of high-redshift galaxies is common to both biased and unbiased samples, it does not affect the result of this comparative analysis. We also stress that the aim of this analysis is not to provide correction factors for our observational results but a way to interpret our results in terms of possible biases introduced by the spectroscopic selection function.

We apply the same procedure to build randomly-extracted mock catalogues with the spectroscopic completeness of the COSMOS field, which has much lower completeness in the group region with respect to the other fields 
(Fig.~\ref{fig:compl_36um}).


\section{Results}
\label{results}

\subsection{The composite groups}
\label{composite_groups}
As already mentioned in the previous sections, in order to follow the evolution of the relation between SF activity and environment, we divide our galaxy sample into four redshift bins, $0<z\leq 0.4$,  $0.4<z\leq 0.8$, $0.8<z\leq 1.2$, $1.2<z\leq 1.7$, according to the redshift distribution of our group sample (see Fig.~\ref{z_histo}). We note that the first three redshift bins comprise 23, 17 and 7 groups, respectively, while the last redshift bin is populated by just one structure at $z\sim 1.6$ \citep{Kurk2009}, which is likely a super-group or a cluster in formation, as suggested by the X-ray analysis (see Section~\ref{s:xray_analysis}). We dedicate a separate section (Sec.~\ref{kurk_sec}) to the discussion about this last redshift bin. 

In each redshift bin, we consider all group galaxies together as members of a composite group. The galaxy group-centric distance (computed from the X-ray centre) in each composite group is normalized to $\rm R_{200}$ of each parent group.  
We then analyse the dependence of the SF activity and stellar mass on the group-centric distance of the composite groups in each redshift bin. This is done to improve the statistics of the group sample. 

To limit the selection effects and, at the same time, to control the different level of spectroscopic completeness per physical properties in the different redshift bins (see e.g. Fig.~\ref{mock_compl_mass_ECDFS}), we apply a common stellar mass cut for the individual galaxies at $\rm 10^{10.3}~M_{\odot}$. This mass cut has three advantages: (a) it corresponds to an IRAC 3.6 $\mu$m apparent magnitude brighter than the nominal 5$\sigma$ detection limit in each considered field up to $z\sim 1.6$; (b) above this limit the spectroscopic completeness is still very high ($>$45\%) in all fields; (c) the considered mass range is still dominated by sources with MIPS and/or PACS detections, thus, with a robust SFR estimate. 
After the mass cut we have 68 galaxies at $0<z\leq 0.4$, 108 at $0.4<z\leq 0.8$, 61 at $0.8<z\leq 1.2$, and 11 sources in the range $1.2<z\leq 1.7$. In all redshift bins the IR detected galaxies are more than 50\%. 

The uncertainties due to the spectroscopic incompleteness of our galaxy sample are evaluated  with dedicated Monte Carlo simulations based on the mock catalogues of \cite{KW_millennium2007} drawn from the Millennium simulation \citep{Springel2005}.
From each of the 100 randomly-extracted mock catalogues (Section~\ref{sec:completeness_mock_cat}) we identify all haloes with masses between $\rm 10^{12.5}$ and $10^{14}~M_{\odot}$ and their members. This information is obtained by linking the mock catalogues of \cite{KW_millennium2007} to the parent halo properties provided by the ``Friend of Friend'' and the \cite{DeLucia2006} semi-analytical model tables of the Millennium data base. 

For any redshift bin used in our analysis, we select a sample of haloes with mass distribution similar to the one of the observed sample of groups. We randomly extract from the group sub-sample in any redshift bin a number of groups equal to that observed. We measure the mean SFR ($\rm{SFR}_{incomplete}$)  as a function of the cluster-centric distance by using the galaxy members of these groups with the same methodology used for the real data set. We repeat this exercise 100 times for each of the 100 randomly-extracted catalogues.  We measure in the same way the mean galaxy SFR ($\rm{SFR}_{real}$) in the original 
\cite{KW_millennium2007} mock catalogues by considering the galaxy members of all groups in each redshift bin with masses in the range $\rm 10^{12.5}-10^{14}~M_{\odot}$. We estimate, then, the difference  ${\rm \Delta SFR=log(SFR_{real})-log(SFR_{incomplete})}$ at the considered values of group-centric distance.  

The dispersion of the distribution of the residual ${\rm \Delta SFR}$ provides the error of our mean SFR at different values of $\rm R/R_{200}$. This error takes into account the bias due to incompleteness, the cosmic variance due to the fact that we are considering small areas of the sky, and the uncertainty in the measure of the mean due to a limited number of galaxies per redshift bin and group-centric distance. The bias introduced by the spectroscopic selection leads to an overestimation of the mean SFR.  This overestimation is independent of the group-centric distance, within the error bars, and is of a factor of 2 in the lowest redshift bin and a factor of 2.5 and 3 in the higher redshift bins. However, we stress here that in less than 0.5\% of the cases this bias leads to a change in the significance of the Spearman correlation test (see next Section). 
We adopt the same procedure for estimating the errors for the other quantities ($\rm M_\star$ and sSFR).

\subsection{SFR as a function of the group-centric distance}
\label{sec:sfr_group-centric_dist}


\begin{figure}
\centering
\includegraphics[width=\hsize]{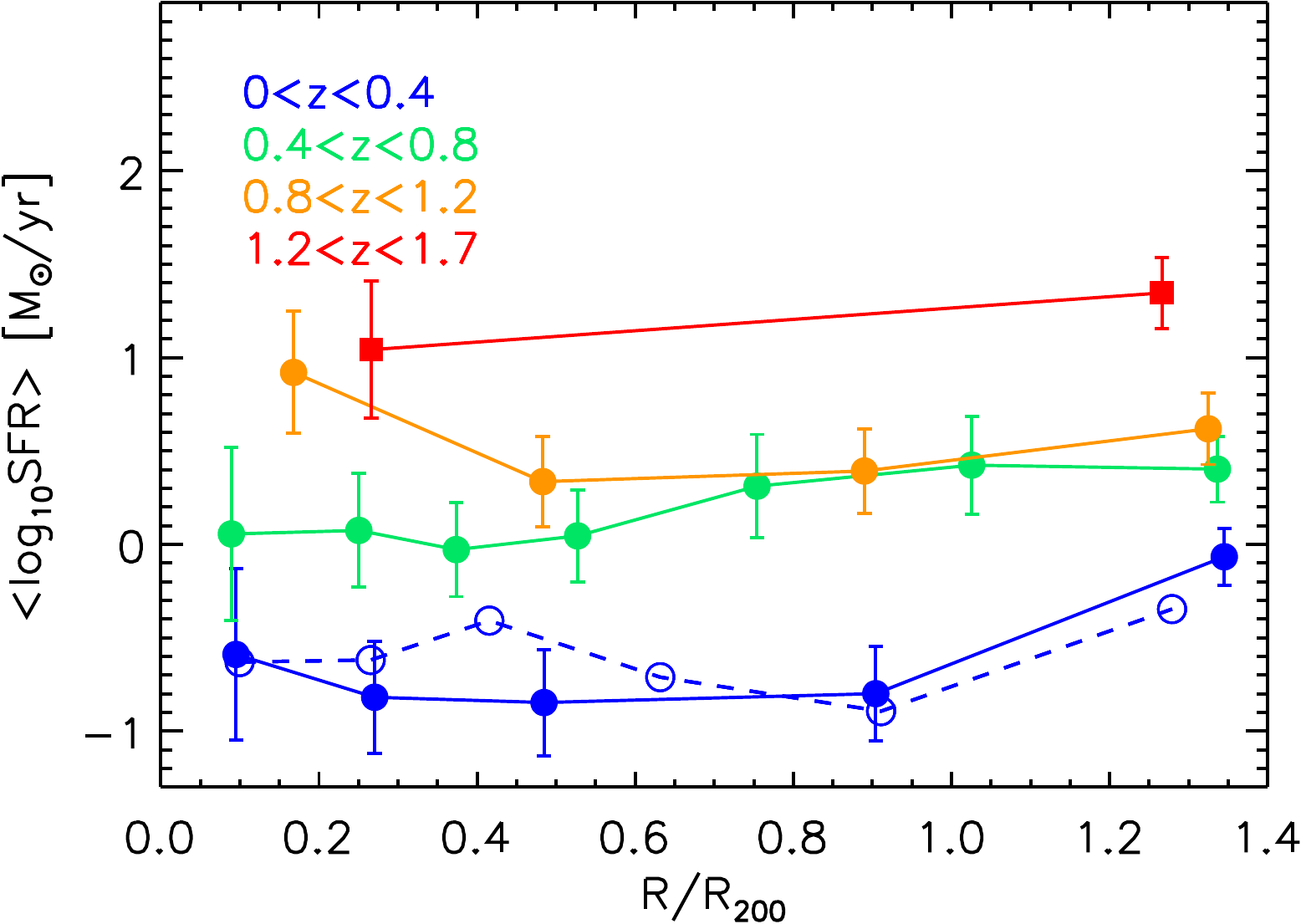}
\caption{Mean SFR as a function of group-centric distance for the composite groups in four different redshift bin. Each point represents the mean SFR among 16 sources up to $z\sim 1.2$. The red squares represent the last redshift bin hosting the super-group at $z\sim 1.6$. They result from the mean SFR among 6 and 5 galaxies. The open symbols connected by a blue dashed line represent the same relation at $0 < z < 0.4$ for all galaxies with $\rm M> 10^9\, M_{\odot}$, with the mean estimated among 30 sources per bin of group-centric distance.  The error bars in Fig.~\ref{fig:sfr_r_plots} are estimated as described in Sec.~\ref{composite_groups}.}
\label{fig:sfr_r_plots}
\end{figure}

We use our data set to shed light on the relation between the mean SFR and the group-centric distance and to follow  its evolution up to $z\sim 1$ with a homogeneous data set. For this purpose we study the mean SFR--group-centric distance relation in the composite groups defined in Sec.~\ref{composite_groups}.
Fig.~\ref{fig:sfr_r_plots} shows our results. We do not find any correlation between $\rm \langle SFR \rangle$ and group-centric distance (as confirmed by the Spearman test) at any redshift. This is consistent with the findings of \cite{Bai2010} who analyse a sub-sample of 9 optically-selected groups at $ 0.06 < z < 0.1$, detected with XMM-$Newton$. Their comparison with rich clusters confirms the different mix of star-forming galaxies within groups compared to clusters.  Our analysis also reveals that the $\rm \langle SFR \rangle$ increases with redshift  (Fig.~\ref{fig:sfr_r_plots}), consistent with the increase in the global SFR out to $z \sim 1$ (\cite{Madau1996} and \cite{Lilly1996}).

\cite{Bai2010} suggest that the continuously decreasing star-forming galaxy fractions towards the centre in the cluster region could reflect a dependence of the SF properties on cluster properties that themselves depend on radius, such as local galaxy density or the density of the intra-cluster medium (ICM). If this is the case, the lack of a dependence of star-forming galaxy fractions on projected radius in groups could be a result of a breakdown of the correlation between galaxy density and projected distance rather than a breakdown of the correlation between star-forming galaxy fractions and galaxy density. To check this issue we analyse also the relation between local galaxy density, estimated similarly to \citet[see also Ziparo et al. submitted for details]{popesso11}, and the group-centric distance in our composite groups. For all of them the Spearman test confirms a clear anti-correlation (significance higher than 5$\sigma$). In contrast, we do not find any relation between the mean SF activity and the density, in agreement with \cite{Peng2012}. Thus, our data confirm a breakdown of the SFR--density anti-correlation within the group regime.

\begin{figure}
\begin{center}
\includegraphics[width=\hsize]{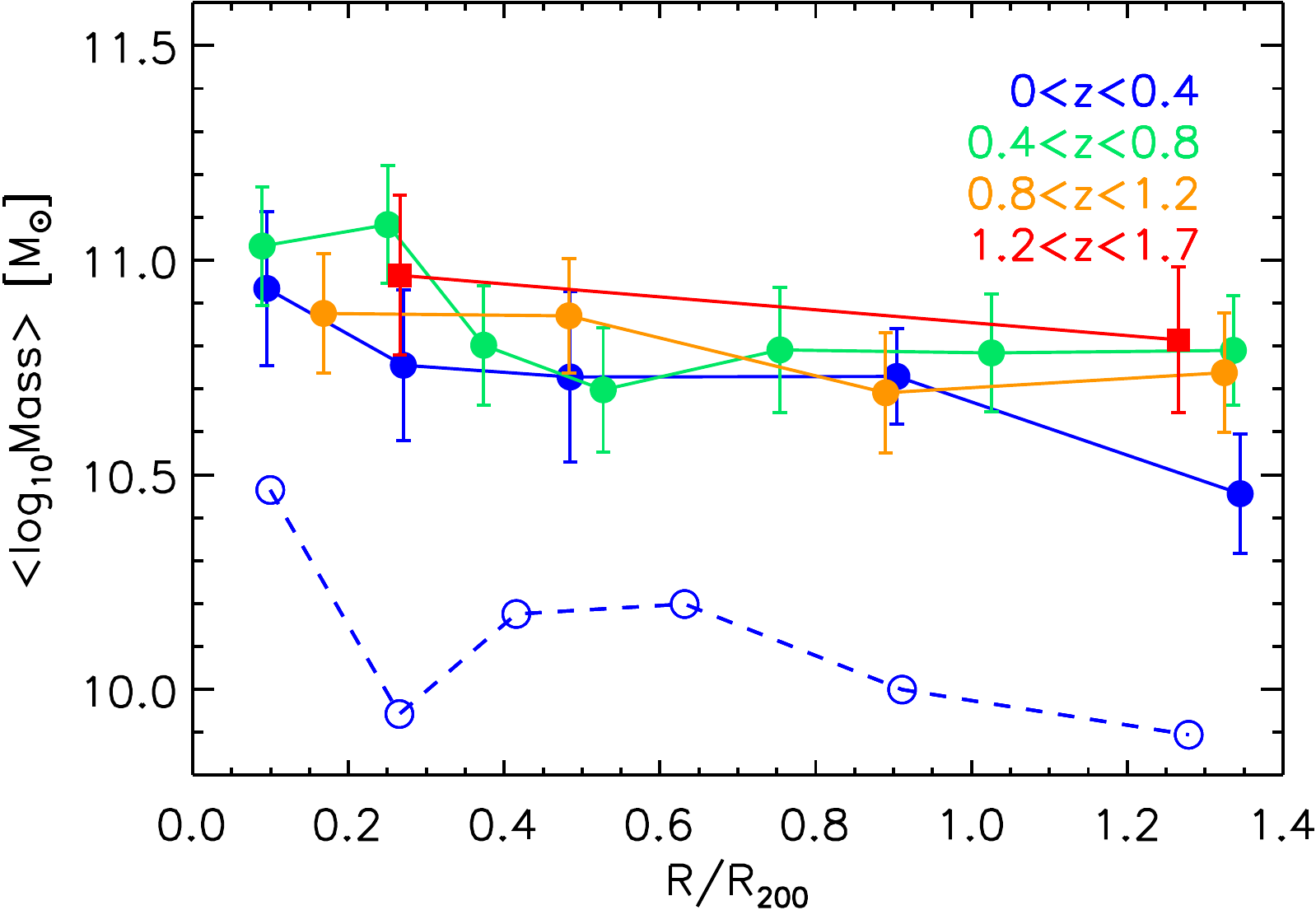}
\end{center}
\caption{Mean galactic stellar mass as a function of group-centric distance for the composite groups in four different redshift bin.  The red squares represent the last redshift bin hosting the super-group at $z\sim 1.6$.  All the galaxies per bin used are the same as in Fig.~\ref{fig:sfr_r_plots}. The open symbols connected by a blue dashed line show the relation between stellar mass and distance from the centre at $0 < z < 0.4$ for all galaxies with $\rm M> 10^9\, M_{\odot}$.  The error bars in Fig.~\ref{fig:sfr_r_plots} are estimated as described in Sec.~\ref{composite_groups}.}
\label{fig:mass_r_plots}
\end{figure}

\subsection{Is there mass segregation in galaxy groups?}

The scenario described in the previous section would be also confirmed by a rather flat relation between the mean stellar mass and the group-centric distance. Indeed, Fig.~\ref{fig:mass_r_plots} shows that there is only a mild ($\sim$ 2.5-3$\sigma$ significance level) anti-correlation between mass and distance from the centre in the two lowest redshift bins and no correlation at all at $z > 0.8$, as confirmed by the Spearman test. The mild correlation in the lowest redshift bins is mainly due to the most central galaxies. Indeed, the mean mass decreases by almost a factor two from the very centre ($\rm R/R_{200}< 0.1$) to $\rm R/R_{200} \sim 0.3-0.4$. However, the large error bars make this difference of low significance. We point out that the analysis of the bias introduced by the spectroscopic selection function conducted with the mock catalogues of the Millennium simulation (see Section \ref{sec:completeness_mock_cat}), shows that a strong degree of mass segregation in the very centre of groups would be hard to observe since massive galaxies with low level of SF activity could be missed by the selection function, which favours massive star-forming galaxies. Thus, given this uncertainty, we cannot exclude that the lack of any level of mass segregation in the analysed groups is caused by a bias introduced by our spectroscopic selection function. 

There could be several different reasons for the lack of strong mass segregation in our group sample. For example, the small mass range considered in our analysis ($\rm M > 10^{10.3} M_{\odot}$) can prevent us from observing a strong underlying mass segregation. To check this possibility, we analyse the stellar mass--group-centric distance relation in the lowest redshift bin with a much lower mass cut of $\rm 10^9 M_{\odot}$. Such an analysis is not possible in the higher redshift bins due to the lower spectroscopic completeness in stellar mass, as shown in Section~\ref{sec:completeness_mock_mass_sfr}. Even after considering lower mass galaxies, we observe only a marginally significant anti-correlation between stellar mass and distance (dashed line in Fig.~\ref{fig:mass_r_plots}). This result is not surprising. Indeed, the presence of strong mass segregation is still a matter of debate even for massive clusters. A classical example is represented by the Coma cluster \citep{White1977} with no significant sign of mass segregation within its virial radius (see also \citealt{Biviano2002} and references therein).

The presence of strong mass segregation is expected to be found in massive clusters as the result of violent relaxation or dynamical friction.  These processes together with progressive accretion, would create in the cluster environment an evolutionary sequence of SF and mass segregation \citep{Gao2004, Weinmann2011}, visible as gradients of mass and SFR. The lack of these gradients in the group environment could indicate that the relaxation or dynamical friction time-scales of group galaxies are longer than the group crossing time and that the spread in accretion times in groups is much smaller than that observed in clusters \citep{Balogh2000}.

Fig.~\ref{fig:mass_r_plots} also shows that the mean stellar mass is rather similar from low to high redshift in agreement with the mild evolution observed for the stellar mass function \cite[e.g.][]{Fontana2004,Fontana2006,ilbertetal2010}.

\begin{figure}
 \begin{center}
\includegraphics[width=\hsize]{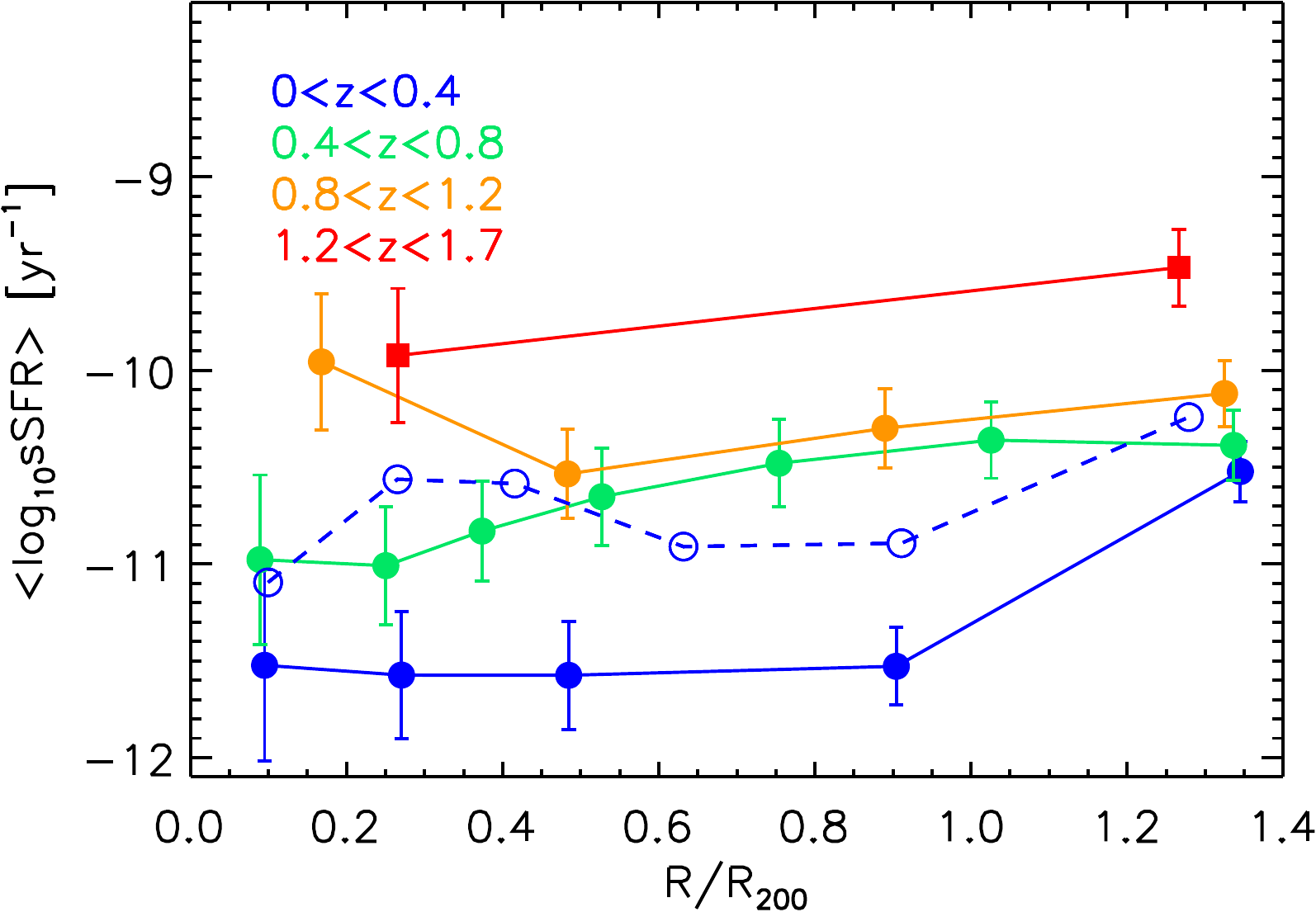}
\end{center}
\caption{Mean sSFR as a function of group-centric distance for the composite groups in four different redshift bins.  The red squares represent the last redshift bin hosting the super-group at $z\sim 1.6$.  All the galaxies per bin used are the same as in Fig.~\ref{fig:sfr_r_plots}. The open symbols connected by a blue dashed line represent the sSFR--group-centric distance relation at $0 < z < 0.4$ for all galaxies with $\rm M> 10^9\, M_{\odot}$.  The error bars in Fig.~\ref{fig:sfr_r_plots} are estimated as described in Sec.~\ref{composite_groups}.}
\label{fig:ssfr_r_plots}
\end{figure}

As a final test we also analyse the mean specific SFR (sSFR)--group-centric distance relation within the group environment (Fig.~\ref{fig:ssfr_r_plots}). As expected due to the lack of strong relation between $\rm \langle SFR \rangle$, $\rm \langle M_\star \rangle$ and group-centric distance, we do not observe any significant relation between $\rm \langle sSFR \rangle$ and the distance from the centre. This still holds for a lower mass cut of $\rm M>10^9~M_\odot$, as we show with a dashed line in Fig.~\ref{fig:ssfr_r_plots}. The error bars of Figs.~\ref{fig:mass_r_plots} and \ref{fig:ssfr_r_plots} are estimated as in Fig.~\ref{fig:sfr_r_plots}, by replacing the SFR with the stellar mass and sSFR in our error analysis.

\section{Discussion}
\label{discussion}

Our results show a lack of gradients in SFR and sSFR and mild mass segregation within X-ray selected groups at all considered redshifts. We discuss in this section the implication of these results, and a comparison with other work. 

\subsection{The absence of star formation gradients}
The weak dependence of $\rm \langle SFR \rangle$ on group global properties, such as the group-centric distance, might be an indication that the SF properties of group galaxies are more affected by their immediate environment, e.g., close neighbours or the presence of substructures \citep{Wilman2005}, than the global environment. However, we have checked that this is not the case. In fact, we find no significant relation between  $\rm \langle SFR \rangle$ and galaxy density. The absence of SF gradient in groups could also suggest that the SF properties of the group members are not directly related to their present environment \citep{Balogh_etal_2004}. This is not usually the case for relaxed clusters, where the local star-forming galaxy fraction increases linearly from the cluster core to large radii in nearby rich clusters \cite[e.g.][]{Balogh2000,Bai+09, Chung+10, Mahajan2010}. 

In particular, \cite{Balogh2000}, in a study of the CNOC1 cluster sample, find that although the  $\rm \langle SFR \rangle$ increases towards the cluster outskirts ($\rm \sim 2 \times R_{200}$), it remains suppressed by almost a factor of two relative to the field. Moreover, the authors reproduce this result by using N--body simulations. Their model assumes that clusters increase their population continuously by accreting field galaxies which are fed by gas from their surroundings. Moreover, as the galaxies enter the cluster potential, reservoirs of fresh fuel for SF are lost. Thus, the origin of radial gradients in these properties is the natural consequence of the strong correlation between radius and accretion time, resulting from the hierarchical assembly of the cluster. 

\begin{figure*}
\begin{center}
\includegraphics[width=0.33\hsize]{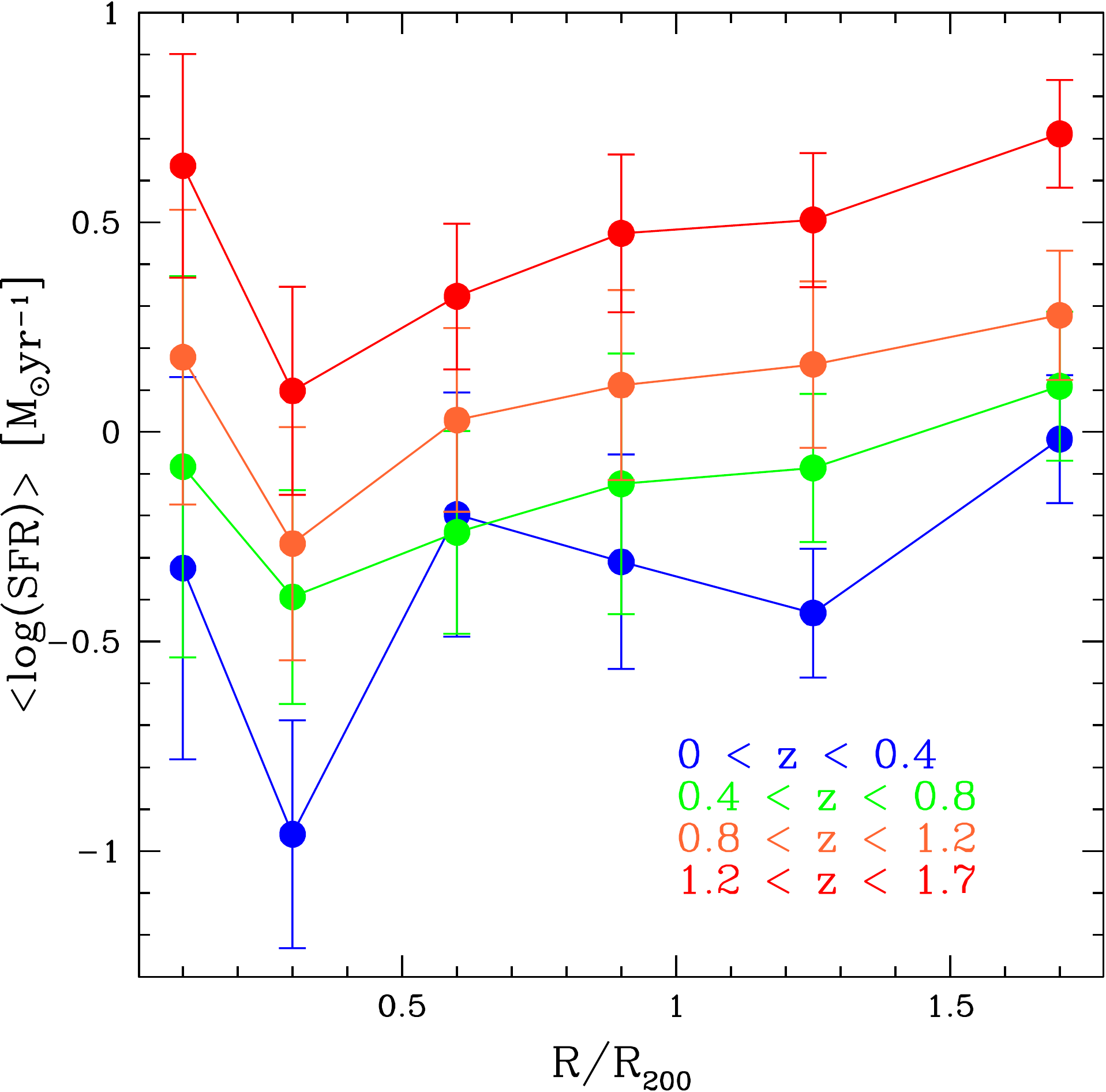}
\includegraphics[width=0.33\hsize]{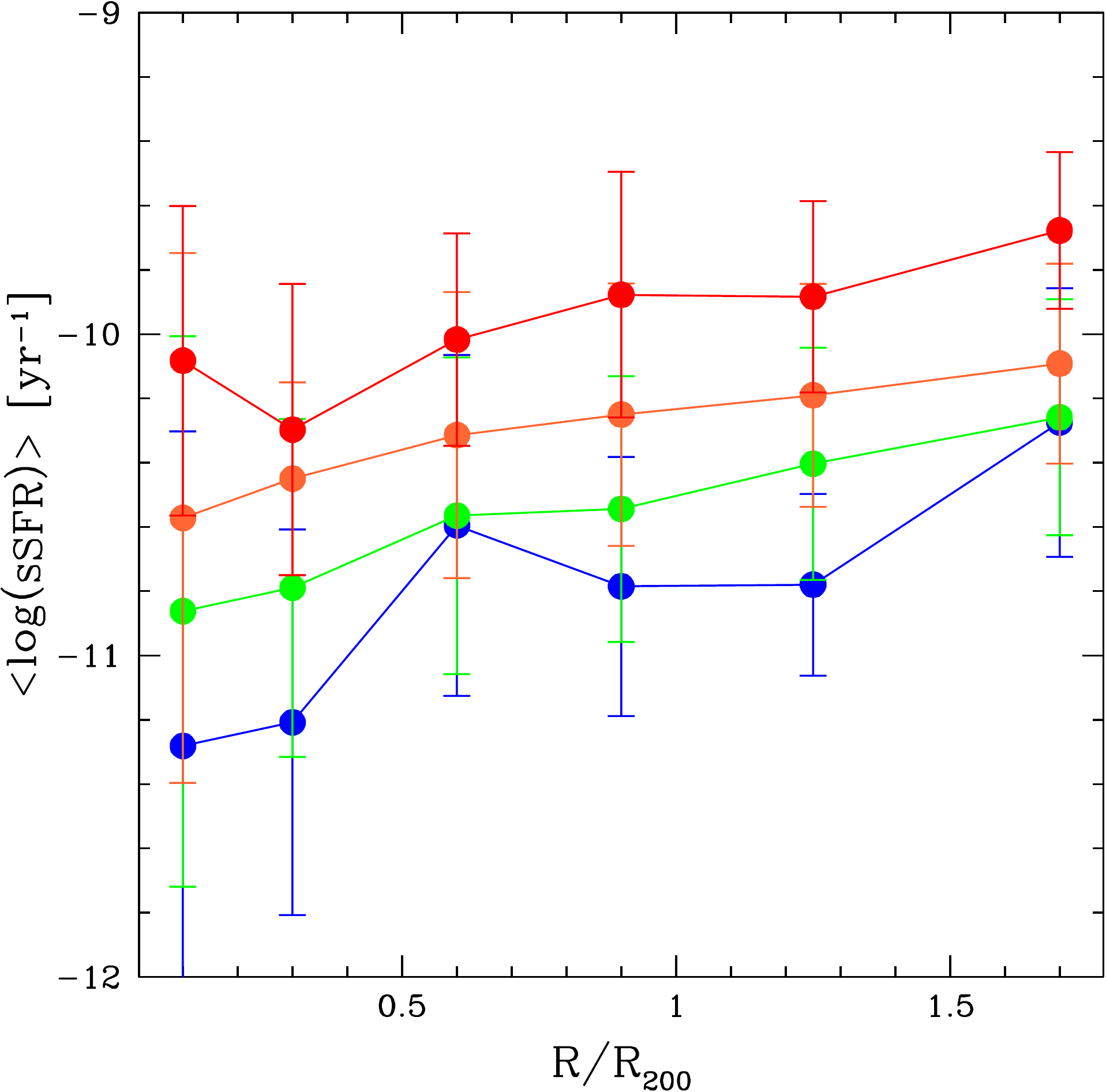}
\includegraphics[width=0.33\hsize]{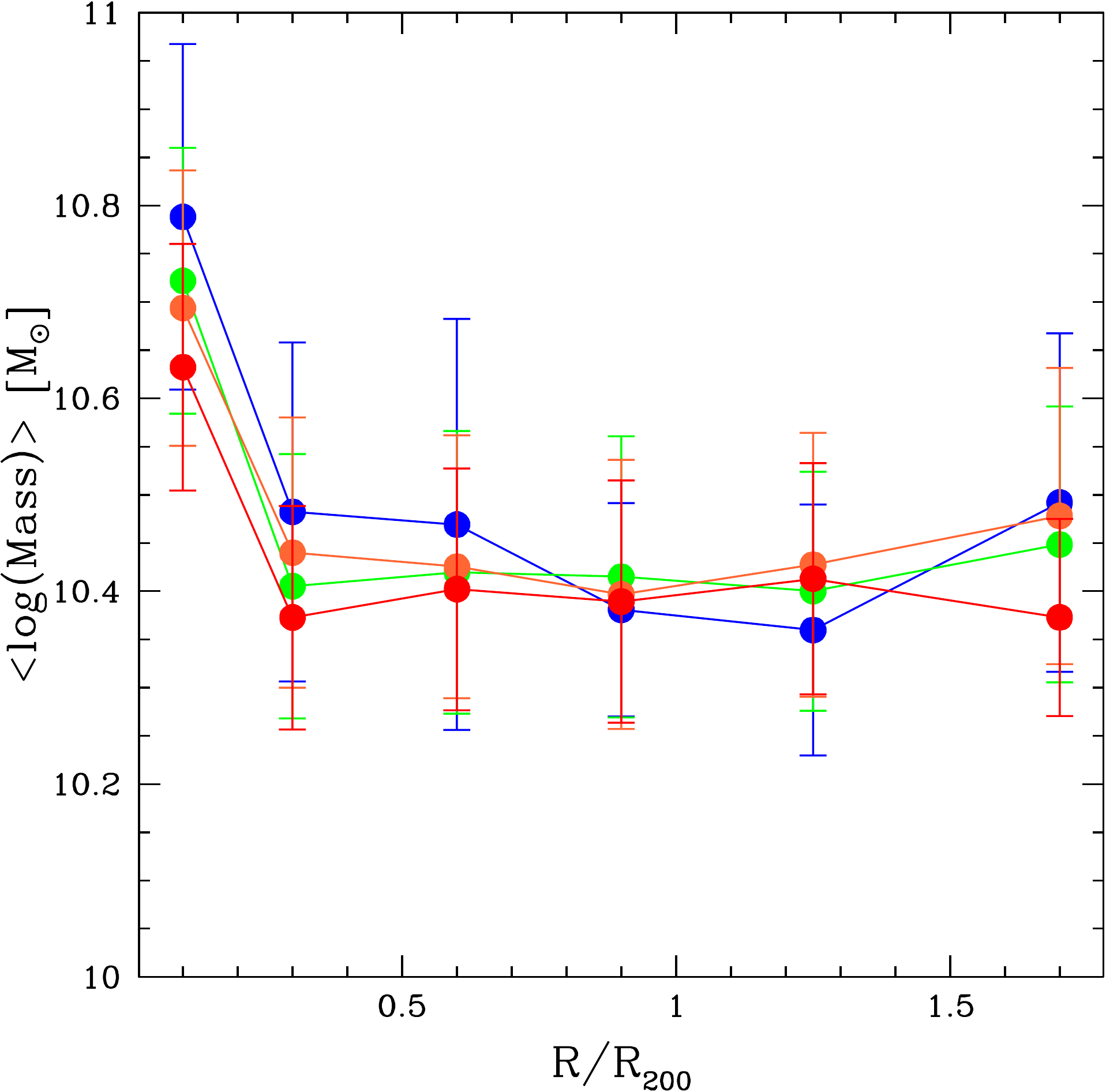}
\end{center}
\caption{Mean SFR (left-hand panel), specific SFR (central panel) and stellar mass (right-hand panel) as a function of the group-centric distance as predicted by the Millennium Simulation mock catalogues of \citet{KW_millennium2007} for group galaxies with $\rm M_\star > 10^{10.3}~M_{\odot}$.}
\label{ms}
\end{figure*}

According to this model, the absence of an anti-correlation between mean galaxy SFR and the group-centric distance could reflect the much smaller spread in accretion times of low-mass objects such as the groups considered in our analysis. This is consistent also with the prediction of the Millennium Simulation. As described in Section~\ref{sec:sfr_group-centric_dist}, we use the mock catalogues of \cite{KW_millennium2007} to build a sample of groups, identified via a friend of friend algorithm \citep{DeLucia2006}, in the same mass range of the observed sample in the different redshift bin. The analysis of the dependence of the SF activity of group galaxies as a function of the group-centric distance shows also that the Millennium simulation does not predict any gradient in SFR or specific SFR, after we apply a mass cut of $\rm M_\star>10^{10.3}~M_\odot$ (Fig. ~\ref{ms}, left and central panel, respectively).
A mild mass segregation is observed only at the very centre (Fig. ~\ref{ms}, right-hand panel). 
Thus, the prediction of mock catalogues is qualitatively in agreement with our observational results. 
The predicted mean SFR is lower than the observed one, in particular at low redshift, due to the satellite overquenching problem \cite[e.g.][]{Gilbank2008, Weinmann2006}. A detailed analysis of the physical reasons of the lack of gradients in the simulation will be carefully discussed in a dedicated paper.

Recent studies have extended this kind of analysis to even larger radii. For example, \cite{Chung+10}, studying a sample of local clusters using WISE data, observe a steep increase in the mean sSFR from the central bin to \rvir, and then an increase until approximately \rvir up to a value below the field value.

Recent results of \cite{Rasmussen2012} and \cite{Wetzel2012} show that a dependence of the SF activity on the distance from the centre is established also in groups. However, both works use optical selection which could introduce some biases in the identification of the groups themselves. In more detail, \cite{Rasmussen2012} analyse a sample of group galaxies at $z\approx 0.06$ with deep UV observations. They detect a SF gradient within $\rm 2 R_{200}$ for galaxies less massive than $\rm 10^{10}~M_\odot$ (similarly to \citealt{Presotto2012}), while they do not find any environmental effect for massive galaxies. The authors argue that the difference in the result with respect to previous works is due to a higher mass cut applied to the other samples. In their opinion, it is in principle possible to observe such a gradient with a higher completeness at low masses. A similar conclusion is reached by \cite{Wetzel2012} who study the fraction of quenched galaxies as a function of group-centric distance in a 
sample of groups in the SDSS. They find that the fraction of quenched galaxies increases towards the halo centre, with a strong trend for the low-mass galaxies.  

According to this scenario, our mass cut at $\rm M=10^{10.3}~M_\odot$ does not allow us to see the ongoing quenching of the SF.  However, analysing the behaviour of galaxies less massive than $\rm 10^{10}~M_\odot$ we do not see any dependence of SFR and sSFR with group-centric distance, even in our lowest redshift bin. The same result is achieved if we apply a mass cut of $\rm M>10^9~M_\odot$, as shown by the blue dashed line in Fig.~\ref{fig:sfr_r_plots}. The divergence might be due to the different range in the group-centric distance considered in the two works. In particular, \cite{Rasmussen2012} consider in their analysis also the group infalling regions (up to $\rm 5\times R_{200}$), where several authors find enhanced SF \cite[e.g.][]{Haines2010, Pereira2010}. On the other hand, our analysis focuses on the study of galaxy properties within $\rm \sim 1.5 \times R_{200}$, thus investigating a region more directly affected by the gravitational potential of the  group.

A conclusion similar to our results is reached by \cite{Bai2010}. As already mentioned, in this work the authors analyse the Spitzer MIPS observations of a sub-sample of 9 groups at $ z\approx 0.06$. These groups are optically selected in the 2dF spectroscopic survey and detected with XMM observations. The authors compare the mean star-forming galaxy (with  $\rm SFR>0.1~M_{\odot}~yr^{-1}$) fraction in the group sample with two clusters using similar data. In contrast to rich clusters, star-forming galaxy fractions in groups show no clear dependence on the distance from the group centres and remain at a level higher than the outer region of rich clusters.  They interpret this result as a possible breakdown of the correlation between the galaxy density and projected distance rather than a breakdown of the correlation between star-forming galaxy fractions and galaxy density. However, we do not find any significant correlation between SF activity and density within the group environment. This  strengthens the interpretation that the  SF properties of the group members are not directly related to their present environment \citep{Balogh_etal_2004}.

\cite{Presotto2012} use a sample of optically selected groups at $0 < z < 0.8$ \citep{Knobel2012} drawn from the $zCOSMOS$ survey \citep{Lillyetal2007}. They observe that the blue fraction of most massive group galaxies ($\rm log(M_{gal}/M_\odot)\geq 10.56$) does not reveal a strong group-centric dependence, even if it is lower than in the field. Conversely, they find a radial dependence in the changing mix of red and blue galaxies for less massive galaxies ($\rm 9.8\leq log(M_{gal}/M_\odot)< 10.56$), with red galaxies being found preferentially in the group centre. They note that this trend is stronger for poorer groups, while it disappears for richer groups. However, their group sample is not X-ray selected. Moreover, \cite{Presotto2012} add photometric members to their groups in order to increase the statistics. This could 
introduce a contamination of field galaxies, in particular in the group outskirts and in poor groups.

In our work, we make use of $Herschel$ PACS data to obtain an accurate estimate of SFR. All the works mentioned above use different SFR indicators which could be affected either by dust obscuration, or by the presence of an AGN. The use of far-IR data, combined with multi-wavelength SED fitting, and checking against spectroscopic selection biases add confidence in our finding about the lack of SF gradients in galaxy groups.

\subsection{The absence of mass segregation}
Our interpretation of the flat relation between mean SFR and group-centric distance is linked to the lack of evidence for mass segregation observed in our groups at any redshift.  Indeed, we find only a mild and low significance ($< 3 \sigma$) anti-correlation between mass and distance from the centre in the two lowest redshift bins and no correlation at all at $z > 0.8$, as assessed by using the Spearman test. 

The lack of mass segregation in groups is observed also in other work in the literature. For example, \cite{Presotto2012} find a  constant mix of galaxy stellar masses irrespective of the radial distance from group centre for poor groups, although they do see significant mass segregation for richer groups.  A similar conclusion is reached by \cite{Tal2012} for a sample of Luminous Red Galaxies (LRGs) at redshift $0.28 < z < 0.4$. Indeed, LRGs are the most massive galaxies ($\rm M>10^{11}~M_\sun$) in the nearby Universe and 90\% of them are expected to be the central galaxy in haloes of $\rm M_{halo}>10^{13}~M_\sun$. Similarly to our result, the authors find a mild mass segregation in LRGs environments (up to 700 kpc from the LRG). It must be noted, however, that \cite{Tal2012} use luminosity segregation to infer mass segregation.  The absence of mass gradient is assessed also 
by \cite{Wetzel2012} who study a sample of optically selected groups in the SDSS. The authors do not find any satellite mass segregation at any group halo mass, which is consistent with our results showing a lack of mass segregation outside the central regions, even at low redshift.

We point out that mass segregation is a matter of debate also in the case of galaxy clusters. Indeed, not all galaxy clusters show a significant sign of mass segregation within the virial radius \cite[e.g.][]{VonDerLinden2010}. \cite{White1977} compare the galaxy distribution observed in the Coma cluster with the one obtained from N--body simulations. The main argument of \cite{White1977} to explain the disagreement between their model and the observation is that most of the cluster mass could not be bound to the galaxies (this is known as ``the missing mass'' problem), since in their model the most massive galaxies end up always in the centre of the cluster (see also \citealt{Biviano2002}). Another plausible explanation can be related to the dynamical state of the cluster. The perturbations due to accretion or merging can delay the relaxation times, since more galaxy encounters are expected.

In general, the presence of strong mass segregation in bound structures is the result of violent relaxation or dynamical friction \citep{Chandrasekhar1943}. In the first case, mass segregation occurs with an exchange of kinetic energy among group galaxies with the lighter galaxies having larger velocity than the heavier galaxies. After the energy is exchanged, most massive galaxies settle in the core of the cluster, while the lighter galaxies preferentially reside in the outer regions. Dynamical friction, instead, represents a kind of frictional drag which causes the galaxy motions to slow down. If a galaxy is in an orbit that makes repeated passages through the cluster or group halo, its orbit will decay over time and it will spiral in and be accreted by the larger object, thus causing that larger object to grow in mass. Since the time-scale of dynamical friction varies as ${\sigma}/{\rho}^3$ (where $\sigma$ is the  velocity dispersion  and $\rho$ the  density of the  halo), high  velocity dispersion clusters do not suffer much internal dynamical evolution of  their galaxy populations after their  primary formation  phase. Conversely, relatively low velocity dispersion  groups could produce interactions and mergers on a cosmologically short time-scale, even  at low redshifts. Thus, if a correlation between the group-centric distance and time since the galaxy infall is expected \citep{Gao2004, Weinmann2011, DeLucia2012}, mass segregation and radial gradients should translate into an evolutionary sequence of SF. However, our results do not support this picture. Instead, they suggest that the  relaxation or the dynamical friction time-scales are too long to lead to a significant mass segregation at any of the redshifts considered.

\subsection{The $z\sim 1.6$ structure}
\label{kurk_sec}
Throughout our analysis we have studied the dependence of galaxy properties on the group-centric distance for a sample of X-ray selected groups. 
We consider also a ``super-group'' or large-scale structure spectroscopically confirmed at $z\sim 1.6$ by  \cite{Kurk2009} and dynamically studied by \cite{popesso2012}. 
This structure allows us to compare at higher refdhift our results on our group sample at $z\lesssim 1$.

The analysis of the CDFS 4 Ms map at the position of the $z\sim 1.6$ structure leads to identification of a few extended X-ray emitting sources possibly associated with this super-group (Finoguenov et al. in preparation).   
One of these extended sources is studied by \cite{Tanaka2012} who report on a relaxed, X-ray bright group, part of the \cite{Kurk2009} structure. 
In this work we study the super group as a large-scale structure rather than single X-ray emitting sources, since the few member identifications to the X-ray emitting sources do not offer good enough statistics to analyse them alone.  The galaxy membership and main structure parameters ($\rm R_{200}$, velocity disperion, $\rm M_{200}$) are derived via dynamical analysis by \cite{popesso2012}.

We must note that this structure could be in the process of formation and, thus, in a particular environmental condition. 
However, the inclusion of this "super-group" does not appear to have a strong influence on our results, as it naturally follows the trends found at $z\sim 1$. 
For instance, Fig.~\ref{fig:sfr_r_plots} shows the mean SFR as a function of system-centric distance. The red curve shows no significant variation between the two radial bins, which represent the mean SFR among 6 and 5 galaxies, respectively. This is confirmed by the Spearman test performed on the 11 galaxies.  

\cite{Tran2010} analyse the dependence of the SF activity as a function of the density in a group at $z=1.6$.  Their $Spitzer$ MIPS data reveals a very high level of SF activity which increases with density. According to their estimate, the highest level of SF happens in the system core. This is apparently in contrast with our results, according to which there is no dependence of SFR on group-centric distance. However, \cite{Tran2010} detect a correlation with a significance of only $2\sigma$. Furthermore, they consider strong IR emitting galaxies whose luminosities could be overestimated due to {extrapolation} of the flux at 24~\um \citep{Elbaz2011}. 
On the other hand, we have used $Herschel$ PACS data for an accurate estimate of SFR. This allows us to avoid contamination by dust obscuration, or by the presence of an AGN. 

Fig.~\ref{fig:mass_r_plots} shows the mean stellar mass as a function of system-centric distance. No significant difference in mean mass can be detected between the two radial bins for the structure at $z\sim 1.6$. The Spearman test confirms the absence of mass segregation in this super-group.  

As for the groups considered in our sample, the absence of a SF gradient and of mass segregation is reflected in the sSFR--system-centric distance relation (Fig.~\ref{fig:ssfr_r_plots}).

Our results are qualitatively consistent with the predictions of the Millennium simulation (Fig. ~\ref{ms}). Indeed, we do not find any gradient of SFR or mass segregation in the simulated groups.

\section{Conclusions}
\label{conclusions}
In the hierarchical scenario of structure formation, galaxy groups are the ``building blocks'' of galaxy clusters. They are also the most common environment of galaxies in the present day Universe, hosting up to 70\% of the galaxy population \citep{Geller_Huchra1983, Eke2005}. Given that most galaxies will experience the group environment during their lifetime, an understanding of groups is critical to follow galaxy evolution in general. 

In order to follow the evolution of the relation between SF activity and the group environment, we have studied a sample of galaxies in X-ray selected groups in four redshift bins, $0<z\leq 0.4$,  $0.4<z\leq 0.8$, $0.8<z\leq 1.2$, $1.2<z\leq 1.7$. To increase the statistics, we have created a composite group in each redshift bin. We note that the last redshift bin is populated by just one structure at $z\sim 1.6$ \citep{Kurk2009}, which is likely a super-group or a cluster in formation.  
With the aim of limiting the selection effects and of taking into account the different level of spectroscopic completeness as a function of each physical property in the different redshift bins, we have applied a stellar mass cut for all galaxies of $\rm M=10^{10.3}$ $M_{\odot}$. 
The uncertainties on the mean galaxy properties are evaluated with dedicated Monte Carlo simulations based on the mock catalogue of \cite{KW_millennium2007} drawn from the Millennium simulation \citep{Springel2005}.

We have analysed the dependence of SF activity on the group-centric distance of the composite groups in each redshift bin. The radial distance from the halo centre is a proxy for the depth of the potential well. We have used our data set to shed light on the relations of the mean mass, SFR and sSFR with the group-centric distance and to follow for the first time their evolution up to $z\sim 1.6$. We find mild mass segregation up to $z\sim0.8$ and no correlation at higher redshift (Fig.~\ref{fig:mass_r_plots}, $\sim2.5-3\sigma$ up to $z\sim0.8$). The mean SFR of galaxies also appears not to be strongly dependent on the distance from the centre (Figs.~\ref{fig:sfr_r_plots} and \ref{fig:ssfr_r_plots}), in contrast to the case for clusters.  Our findings are in agreement with the predictions of the Millennium Simulation mock catalogues of \cite{KW_millennium2007}, as we show in Fig.~\ref{ms}. 
The absence of any measurable segregation of SF activity within the group environment could reflect the much smaller spread in accretion times of groups with respect to clusters. This result is not 
affected by the mild mass segregation that we observe up to $z\sim0.8$, which indicates that the relaxation or the dynamical friction time-scales of group galaxies is longer than the group crossing time. Thus, the time necessary for heavy galaxies to sink to the centre of the potential well is too long compared to the lifetime of the group itself.

\section*{Acknowledgements}
We thank the anonymous referee for her/his constructive comments.

FZ acknowledges the support from and participation in the International Max-Planck Research School on Astrophysics at the Ludwig-Maximilians University.

MT gratefully acknowledges support by KAKENHI No. 23740144.

FEB acknowledges support from Basal-CATA (PFB-06/2007), CONICYT-Chile (under grants FONDECYT 1101024, ALMA-CONICYT 31100004 and Anillo ACT1101) and {\it Chandra} X-ray Center grant SAO SP1-12007B.

PACS has been developed by a consortium of institutes led by MPE 
(Germany) and including UVIE (Austria); KUL, CSL, IMEC (Belgium); CEA, 
OAMP (France); MPIA (Germany); IFSI, OAP/AOT, OAA/CAISMI, LENS, SISSA 
(Italy); IAC (Spain). This development has been supported by the funding 
agencies BMVIT (Austria), ESA-PRODEX (Belgium), CEA/CNES (France),
DLR (Germany), ASI (Italy) and CICYT/MCYT (Spain).

We gratefully acknowledge the contributions of the entire COSMOS
collaboration consisting of more than 100 scientists. More information
about the COSMOS survey is available at
http://www.astro.caltech.edu/$\sim$cosmos.

This research has made use of NASA's Astrophysics Data System, of NED,
which is operated by JPL/Caltech, under contract with NASA, and of
SDSS, which has been funded by the Sloan Foundation, NSF, the US
Department of Energy, NASA, the Japanese Monbukagakusho, the Max
Planck Society and the Higher Education Funding Council of England.
The SDSS is managed by the participating institutions
(www.sdss.org/collaboration/credits.html).

This work has been partially supported by a SAO grant SP1-12006B grant to UMBC.

\bsp
\label{lastpage}

\end{document}